\documentclass[acmsmall,screen,nonacm]{acmart}
\pdfoutput=1
\acmJournal{PACMPL}
\acmVolume{1}
\acmNumber{PLDI}
\acmArticle{1}  
\acmYear{2024}
\acmMonth{6}
\acmDOI{}
\startPage{1}

\setcopyright{none}

\usepackage{booktabs}
\usepackage{adjustbox}
\usepackage{algorithm}
\usepackage{algpseudocode}
\usepackage{mathtools}
\usepackage[nounderscore]{syntax}
\usepackage{tikz}
\usetikzlibrary{shapes.geometric,shapes.misc,shapes.multipart,positioning,fit,matrix,patterns,calc,arrows.meta}
\usepackage{pgfplots}
\usepackage{wrapfig}
\pgfplotsset{compat=1.18}
\tikzset{every picture/.style={font={\rmfamily}}}
\usepackage{cleveref}
\usepackage{pifont}
\usepackage{thmtools}

\definecolor{cat01}{HTML}{1f77b4}
\definecolor{cat02}{HTML}{ff7f0e}
\definecolor{cat03}{HTML}{2ca02c}
\definecolor{cat04}{HTML}{d62728}
\definecolor{cat05}{HTML}{9467bd}
\definecolor{cat06}{HTML}{8c564b}
\definecolor{cat07}{HTML}{e377c2}
\definecolor{cat08}{HTML}{7f7f7f}
\definecolor{cat09}{HTML}{bcbd22}
\definecolor{cat10}{HTML}{17becf}

\definecolor{lsqborange}{RGB}{237,183,112}
\definecolor{lsqbblue}{RGB}{144,177,207}
\definecolor{lsqbred}{RGB}{227,139,120}
\definecolor{lsqbpink}{RGB}{240,209,227}
\definecolor{lsqbpurple}{RGB}{189,188,215}

\algdef{SxnE}[ALIRFOR]{ALIRFor}{EndALIRFor}[1]{\algorithmicfor\ #1}
\algdef{SxnE}[ALIRLET]{ALIRLet}{EndALIRLet}[1]{\textbf{let}\ #1\ $\coloneqq$}
\algdef{cxnE}{ALIRLET}{ALIRIn}{EndALIRLet}
\algrenewtext[ALIRLET]{ALIRIn}{\textbf{in}}

\algblockdefx[COMPACTIF]{CompactIf}{CompactElseEndIf}[1]{\algorithmicif\ #1\ \algorithmicthen}[1]{\algorithmicelse\ #1\ \algorithmicend\ \algorithmicif}

\newcommand*{\lB}{\{\mskip-5mu\{}
\newcommand*{\rB}{\}\mskip-5mu\}}

\DeclareMathOperator{\supp}{supp}
\newcommand*{\Bool}{\mathrm{Bool}}
\newcommand*{\dblequals}{\mathrel{\mathrlap{\raisebox{0.1em}{-}}\raisebox{-0.1em}{-}\mathrlap{\raisebox{0.1em}{-}}\raisebox{-0.1em}{-}}}
\DeclareMathOperator{\floor}{floor}
\newcommand*{\interp}{\mathcal{I}}
\newcommand*{\gallopV}{\mathcal{G}}

\newcommand*{\tabyes}{\Pisymbol{pzd}{51}}
\newcommand*{\tabno}{\Pisymbol{pzd}{55}}

\newcommand*{\agg}{\mathsf{G}}

\newcommand{\hide}[1]{}

\newcommand*{\vennThreeFirst}{(90:0.875cm) circle (1.25cm)}
\newcommand*{\vennThreeSecond}{(210:0.875cm) circle (1.25cm)}
\newcommand*{\vennThreeThird}{(330:0.875cm) circle (1.25cm)}

\newcommand*{\plusequals}{\mathrel{\mathord{+}\mathord{=}}}

\newcommand*{\ojoinbars}{\rule[0.05ex]{.25em}{.6pt}\llap{\rule[1.0ex]{.25em}{.6pt}}}
\DeclareMathOperator*{\ljoin}{\ojoinbars\mkern-6.8mu\Join}
\DeclareMathOperator*{\rjoin}{\Join\mkern-6.8mu\ojoinbars}
\DeclareMathOperator*{\ojoin}{\ojoinbars\mkern-6.8mu\Join\mkern-6.8mu\ojoinbars}
\DeclareMathOperator*{\ijoin}{\Join}

\newcommand*{\NULL}{\textsf{NULL}}
\newcommand*{\dups}{\textsf{dups}}

\title{A Compiler for Operations on Relations with Bag Semantics}

\author{James Dong}
\orcid{0000-0001-9281-4156}
\affiliation{%
  \institution{Stanford University}
  \streetaddress{450 Jane Stanford Way}
  \city{Stanford}
  \state{CA}
  \postcode{94305}
  \country{USA}}
\author{Fredrik Kjolstad}
\orcid{0000-0002-2267-903X}
\affiliation{%
  \institution{Stanford University}
  \streetaddress{450 Jane Stanford Way}
  \city{Stanford}
  \state{CA}
  \postcode{94305}
  \country{USA}}

\begin{abstract}
    We describe an abstract loop-based intermediate representation that can express fused implementations of relational algebra expressions on sets and bags (multisets). The loops are abstracted away from physical data structures thus making it easier to generate, reason about, and perform optimization like fusion on. The IR supports the natural relational algebra as well as complex operators that are used in production database systems, including outer joins, non-equi joins, and differences. We then show how to compile this IR to efficient C++ code that co-iterates over the physical data structures present in the relational algebra expression. Our approach lets us express fusion across disparate operators, leading to a 3.87$\times$ speedup (0.77--12.23$\times$) on selected LSQB benchmarks and worst-case optimal triangle queries. We also demonstrate that our compiler generates code of high quality: it has similar sequential performance to Hyper on TPC-H with a 1.00$\times$ speedup (0.38--4.34$\times$) and competitive parallel performance with a 0.61$\times$ speedup (0.23--1.80$\times$). Finally, our approach is portable across data structures.
\end{abstract}

\begin{document}
\maketitle

\section{Introduction}
\label{sec:introduction}

Database management systems lower SQL or dataframe operations to a relational algebra IR consisting of operators like joins, filters, and aggregation. A logical query planner optimizes the relational algebra, e.g., to push operations that filter out work before other operations~\cite{architecture-db,overview-query-opt}. A physical query planner then selects appropriate implementations for different operations in the relational algebra from a library of hand-written implementations, such as a sort-merge joins, hash joins, and filter implementations~\cite{dbs-complete-book}.

By selecting a fixed number of hand-written implementations, we leave performance on the table, as there is a limit to the number of variants implementers can write. In practice, this limit means that we cannot fuse multiple operators together. Fused operators yield better temporal locality and, as the literature on worst-case optimal joins shows~\cite{ngo-skew}, can provide asymptotically better performance. The lack of fusion can therefore have serious performance consequences.

However, the irregular nature of relational algebra makes it challenging to fuse an arbitrary combination of relational operators. These operators are data dependent, there are many different types of joins (e.g., outer and non-equi joins) and join implementations (e.g., sort-merge and hash), and the relations are stored in diverse data structures such as column stores, hash maps, tries, and B-trees. Fused code must therefore co-iterate over any number of irregular data structures to implement a large set of joins and other operations. Finally, practical databases use bag (multiset) semantics~\cite{on-multisets-in-dbs,multiset-ra,multiset-table-algebra}, complicating execution.

Prior work has shown how to generate fused code for some combinations of some relational operators. The line of work that began with the Hyper system~\cite{hyper} showed how to fuse operator pipelines, which are straight-line sequences of operators, by composing user-written C++ template implementations of operators. The EmptyHeaded system~\cite{emptyheaded} took fusion one step further and allowed inner joins to be fused together, which prior work has shown is necessary to achieve worst-case optimal performance on algorithms like triangle queries. Finally, prior work on indexed streams~\cite{indexed-streams} showed how to fuse the operators in the natural relational algebra on sets, but not outer joins or non-equi joins, and not on multisets. In conclusion, these systems are limited in what kinds of operators they can fuse, as shown in \Cref{tab:comparison}.

\begin{table}[tbp]
    \centering
    \scriptsize
    \caption{Comparison of different approaches to relational algebra compilation.}
    \vspace{-1em}
    \begin{tabular}{lcccc}\toprule
        System & Push/Pull & Fusion & \shortstack{Bag Semantics, \\ Outer Join} & \shortstack{Data Structure \\ Portability} \\\midrule
        Volcano & Pull & None & \tabyes & \tabyes \\
        Hyper & Push & Single Pipeline & \tabyes & Limited \\
        EmptyHeaded & Push & Multi-way Inner Join & \tabyes & Limited \\
        Indexed Streams & Push & Natural RA (sets) & \tabno & \tabyes \\
        Ours & Push & Full RA & \tabyes & \tabyes
    \\\bottomrule
    \end{tabular}
    \label{tab:comparison}
\end{table}

We describe an approach that can be used to generate a fused implementation for any set of relational algebra operators that can be fused, including projections, filters, inner joins, outer joins, left/right joins, differences, and non-equi joins. Our approach can fuse operators on both natural set relations and the relations with multiset/bag semantics that are ubiquitous in practical database systems. Our approach lowers a graph of relational algebra operators to a novel loop-based IR. The loops in the IR iterate over a data model that is abstracted away from concrete data structures. Our technical contributions are:
\begin{enumerate}
    \item A data model based on coordinate trees that is abstracted away from concrete data structures, thus providing portability across data structure. The data model supports both set semantics and bag semantics (\Cref{sec:data-model}).
    \item An intermediate representation (ALIR) that can represent fused relational algebra operators as loops that hierarchically iterates over the coordinate tree data model (\Cref{sec:abstract-loops}). 
    \item \emph{Iteration machines} that represent loop domains of arbitrary multiset expressions, for generating low-level co-iteration code for loops over relational columns (\Cref{sec:iteration-machines}).
    \item A code generation algorithm from the abstract loops to low-level C++ code (\Cref{sec:code-generation}).
\end{enumerate}
We show that our approach can generate efficient code without sacrificing generality or data structure portability: compared to Hyper~\cite{hyper} the performance of our generated code on TPC-H is on (geomean) average 1.00$\times$ (0.38--4.34$\times$) compared to the Hyper system in sequential execution, and 0.61$\times$ (0.23--1.80$\times$) in parallel execution. However, the benefit of our system is its ability to generate fused code. On four LSQB graph query benchmarks, which provide opportunities for fusion across different types of operators, we achieve an average speedup of 3.87$\times$ in parallel (0.77--12.23$\times$).

\section{Overview}
\label{sec:overview}

\begin{figure}[b!]
    \footnotesize
    \vspace{-1em}
    \begin{tikzpicture}[node distance=0.7cm]
        \node (ra) {\shortstack{Relational\\Algebra}};
        \node (alir) [draw,right=of ra,fill=gray,fill opacity=0.1,text opacity=1] {ALIR};
        \node (sd) [below=0.3cm of ra] {\shortstack{Storage\\Description}};
        \node (mid) [coordinate] at ($(ra)!0.5!(sd)$) {};
        \node (mid2) [coordinate] at (alir|-mid) {};
        \node (out) [rounded corners,row sep=0.3em,draw,matrix,align=center,right=of mid2,
        fill=gray, fill opacity=0.1,text opacity=1
        ] {
            \node[anchor=north,draw,rounded corners=0,
            fill=gray, fill opacity=0.2,text opacity=1
            ]{\shortstack{Iteration\\Machines}}; \\
            \node[anchor=north,]{\shortstack{Code\\Generation}}; \\
        };
        \node (code) [right=of out] {C++ Code};
        \draw[->,>=stealth] (ra) -- (alir);
        \draw[->,>=stealth] (alir) -- (out.west|-alir);
        \draw[->,>=stealth] (sd) -- (out.west|-sd);
        \draw[->,>=stealth] (out) -- (code);
    \end{tikzpicture}
    \caption{Flowchart of our compiler. 
    Square boxes show intermediate representations.}
    \Description{A flowchart, with relational algebra translated into ALIR, then converted into iteration machines during code generation, to which storage description is an additional input, finally producing C++ code.}
    \label{fig:flowchart}
    \vspace{-1em}
\end{figure}
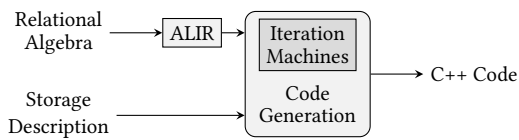

Our approach produces code from the following inputs:
the expression to be generated, converted to abstract loop IR (ALIR),
which captures the iteration pattern of the target expression, and
the storage description, a description of how the
input is structured.
Depending on various factors such as the distribution of values in
the dataset, desired performance characteristics, and data layout,
a relational expression has several distinct versions of
ALIR which have equivalent semantics but differing execution
strategies. As this is not the focus of the
paper, we perform this conversion semi-automatically and manually fine-tune it to
optimize performance on the benchmark data.

The overall structure of our approach is shown in
Figure~\ref{fig:flowchart}.
Given an expression in ALIR, we break the iteration structure into
nested loops, each over a particular single-dimensional domain.
For each domain, we produce an \emph{iteration machine},
which expresses both how iteration proceeds as input relations are
exhausted, as well as how to handle the cases within each loop,
depending on which relations are present.
As they are being generated, we automatically perform
various optimizations on the iteration machines to improve performance
and reduce the amount of generated code.

Finally, we lower the code from ALIR with iteration machines to
native C++ code. An additional input to this stage is the storage
description,
which specifies the possible ways that data should be accessed. For
example, data which is stored as a hash table cannot be accessed
sequentially in order, only via lookup for a given key. Conversely,
data stored as a sorted list can be iterated in order, but looking up
a particular key is not very efficient when co-iteration is possible.

Given the iteration machines and storage description for each relation,
our approach lowers code by generating loops that
simultaneously iterate (co-iterate) over input relations.
These loops are fused, meaning
they do not generate any intermediate results in auxiliary storage,
even for arbitrarily complex expressions, although the user may choose to precompute values.

\section{Data Model}
\label{sec:data-model}

How relations are stored is an important and complex part of efficiently performing relational algebra computations~\cite{access-path,column-row}.
In the simplest case, each tuple in a relation is stored as a single object in a list in memory.
However, performant databases use several methods, such as storing certain types of data in separate storage or
auxiliary index data structures to improve performance.
In order to effectively fuse arbitrarily complex operations, we need a data model that allows us to
break down accesses into per-attribute granularity.
As such, each attribute (dimension) of a relation has its own data storage to allow for maximum flexibility,
but we abstract away the implementation details of how data is accessed and use general interfaces to perform data access.
These dimensions are linked together in a structure known as a \emph{coordinate tree.}

\subsection{Relations as Coordinate Trees}
\begin{wrapfigure}{l}{0.31\textwidth}
    \footnotesize
    \begin{tikzpicture}[
        node/.style={circle,draw,minimum width=2em},
        x=1.5cm,y=-1.75cm,
        scale=0.5
    ]
        \node (root) at (0,0) [node] {};
        \node (a1)   at (0,1) [node] {1};
        \node (a2)   at (2,1) [node] {2};
        \node (b1)   at (0,2) [node] {1};
        \node (b2)   at (1,2) [node] {3};
        \node (b3)   at (2,2) [node] {2};
        \node (c1)   at (0,3)        {(1, 1)};
        \node (c2)   at (1,3)        {(1, 3)};
        \node (c3)   at (2,3)        {(2, 2)};
        \node (c4)   at (3,3)        {(2, 2)};
        \node at (4.5,0) {level 0};
        \node at (4.5,1) {level 1};
        \node at (4.5,2) {level 2};
        \node at (4.5,3) {level 3};
        \draw (root) edge (a1) edge (a2);
        \draw (a1) edge (b1) edge (b2);
        \draw (a2) edge (b3);
        \draw[dashed] (b1) edge (c1);
        \draw[dashed] (b2) edge (c2);
        \draw[dashed] (b3) edge (c3) edge (c4);
    \end{tikzpicture}\vspace{-1em}
    \caption{Example coordinate tree. Nodes are labeled with
    their values; the tuples in each subtree is the the multiset for each node.
    Dotted lines represent the tuples corresponding to each
    leaf node.}\vspace{-1em}
    \label{fig:coordtree}
\end{wrapfigure}
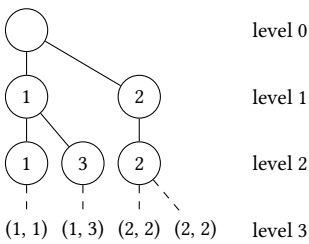
Unlike standard relational algebra, in which relations are sets,
we model relations using \emph{bag} or \emph{multiset} semantics, like most real-world RDBMSes.
In this view, a relation is taken as a \emph{multiset} of tuples, or
rows, of fixed length, where each element in these tuples is called an
attribute, or column. For convenience, when talking about the
attributes of a relation, we assign each attribute a
distinct name. For example, a relation with two attributes named $a$
and $b$ would have tuples of the form $(a, b)$, and we use the
notation $\tau.a$ to extract the value of attribute $a$ from a tuple $\tau$.
For a relation $R$, we use the notation $R.a^*
\coloneqq \lB \tau.a \mid \tau \in R \rB$ to represent the values of particular attribute, and $R.a = \supp(R.a^*)$
(where the double brackets distinguish multisets from regular sets,
and $\supp(M) = \{ x \mid x \in M \}$ is the set of elements in a multiset.)

Our approach represents a relation $R$ with attributes $a_1, \dots, a_r$
as a \emph{coordinate tree}~\cite{kjolstad2017}
consisting of one level for each attribute, in some predefined order.
Although the physical representation of the data may differ, we use coordinate trees as an
abstract representation to model data.
In a coordinate tree,
attributes are stored hierarchically, where level 0 represents all of $R$,
and the children at level $i$ represent the distinct values of attribute $a_i$
within the multiset of elements that match all previous attributes
$\lB \tau \in R \mid \bigwedge_{j < i} \tau.a_j = v_j \rB$, where $v_j$ is the value of the
ancestor node at level $j$.
In this representation, each element of $R$ is specified by the values of a path from the root to
a leaf, and there is exactly one such path for each element in $R$.

\subsection{Relation Storage}
The storage description for a relation consists of layer storage for
each layer in the coordinate tree.
Layer storage describes how the data in each level can be accessed.
For example, a layer may be stored as a sorted list of values along
with a reference to the storage for the next level, a hash table
from attribute value to row indices, or a dense set over all possible
attribute values. This is done by a small set of primitive routines,
defined for each layer storage implementation.

These routines are grouped into two \emph{capabilities}: lookup and iteration,
and which capabilities a layer has influences how co-iteration is performed for a given set of layers.
Co-iteration refers to iterating over all values in multiple inputs (here layers) simultaneously.
For each set of layers, co-iteration is performed only over a subset with iteration capability,
and the remaining layers must have lookup capability. The choice of which layers use which capability
depends both on the desired performance characteristics and the structure of the computation,
as not every choice of layer capabilities can be used to generate correct code (see \Cref{sec:legality}).

\paragraph{Lookup.} Layers with lookup capability have the interface of a membership query.
The following routines
are required a layer of type $T$ to have lookup capability:

\begin{itemize}
    \item $\texttt{init} \colon () \longrightarrow ()$.
    This performs any necessary initialization for the layer.
    \item $\texttt{present} \colon T \longrightarrow \Bool$. This returns whether or not a given value is present in a layer.
\end{itemize}

\paragraph{Iteration.} Layers with iteration capability (iterators) have the interface of an iterator
\cite{indexed-streams}
over the set of layer's values,
which is required to be sorted if there are any other iterator layers participating in co-iteration.
Iteration capability requires the following routines:

\begin{itemize}
    \item $\texttt{init} \colon () \longrightarrow ()$. This initializes the iterator at the beginning.
    \item $\texttt{curval} \colon () \longrightarrow T$. This returns the value of the current element.
    \item $\texttt{advance} \colon () \longrightarrow ()$. This advances the iterator to the next element.
    \item $\texttt{present} \colon T \longrightarrow \Bool$,
        defined as\, $\texttt{present}(v) \coloneqq \mathbf{let}\ c \gets \texttt{curval}()\ \mathbf{in}\ c \dblequals v$.
        This has the same function as for lookup capability, but it is sufficient to only check the current element.
    \item $\texttt{skipto} \colon T \longrightarrow ()$. This advances the iterator to the first element greater than
    or equal to the given value, which for many data types is more efficient than repeatedly calling \texttt{advance}.
    \item $\texttt{reset} \colon () \longrightarrow ()$ (required only in Cartesian products). This resets the iterator to the beginning.
\end{itemize}

This framework for describing relation storage suffices to
represent many common storage types, such as 
    row and
    column storage,
    tries,
    hash and B-tree indices, as well as dense and sparse matrices.

\subsection{Related Values and Primary Keys}
In many relations, a small subset of attributes form a \emph{primary key}, and the other attributes, which we refer to as \emph{related values}, are not typically iterated over but instead treated as additional data to be looked up~\cite{dbs-complete-book}. The values in the primary key are known to be unique, so the related values form a single chain under the primary key attributes. Due to this structure, it is not necessary to iterate over these attributes. Instead, related values are attached to the storage for primary keys, and the primary key storage is responsible for maintaining iteration state or looking up values as necessary.
We can use the same abstraction to model data structures such as tensors and arrays, by treating the coordinates as the primary key and the values as related values.

\section{Abstract Loop IR}
\label{sec:abstract-loops}

\begin{figure}
    \begin{minipage}{0.4\textwidth}
    \begin{grammar}\footnotesize
        <stmt> ::= \textbf{for} $\alpha \in $ <multiset-expr> <stmt>\dots
            \alt ( <expr>, \dots )
            \alt R [ <expr>, \dots ].$\alpha$ <assign-op> <expr> (\ref{subsec:aggregation})
            \alt \textbf{let} $R(\alpha, \dots)$ \textbf{in} <stmt>\dots\ (\ref{subsec:precomputation})
        
        <expr> ::= $\alpha$ | (constant) | <expr> $\circ$ <expr> | \dots

        <assign-op> ::= $=$ | $\plusequals$ | \dots
    \end{grammar}
    \end{minipage}
    \begin{minipage}{0.5\textwidth}
    \begin{grammar}\footnotesize
        <multiset-expr> ::= $R.\alpha$
            \alt \{ $\alpha \mid $ <expr> \}\quad |\quad \{ <expr> \} (\ref{subsec:virtual-inputs})
            \alt <multiset-expr> <multiset-op> <multiset-expr>
            \alt <multiset-expr> (<multiset-expr>) (\ref{subsubsec:lookup-only-relations})
            \alt <multiset-expr> $\cup\ \varnothing$ (\ref{subsec:outer-joins})

        <multiset-op> ::= $\cup$ | $\cap$ | $-$ | $+$ | $\times$
    \end{grammar}
    \end{minipage}
    \vspace{-0.5em}
    \caption{Grammar for ALIR. $R$ represents a relation, and $\alpha$ is an attribute.}
    \Description{Grammar for ALIR, including update syntax for aggregations, let/in for precomputation, singleton relations, index-only relations, and $\cup \varnothing$ syntax.}
    \label{fig:alir-grammar}
    \vspace{-1em}
\end{figure}

While relational algebra is capable of expressing many types of computation across several domains,
it is not very suitable for generating fused code, as naively computing each relational operation
results in many unneeded temporary relations.
This is true for any representation for which intermediate results are represented as relations.
A representation such as an operator graph that relies on templating can only handle optimizations
with a fixed structure for each template, so such representations are
too coarse-grained to model fusion across arbitrary relational expressions.

Instead, we use a representation that focuses on \emph{fusion by default}, which represents the
computation as a series of nested loops.
That is, computation is split the same way as the data is broken up, with nested loops iterating over separate attributes.
The flexibility of this approach allows us to model arbitrary fused expressions.
We call our notation the Abstract Loop Intermediate Representation, or ALIR, with the grammar in Figure~\ref{fig:alir-grammar}. In the sections that follow, we describe different concepts in ALIR as well as the relational algebra operations for which they are used.

\begin{wrapfigure}{r}{0.23\textwidth}
\vspace{-1.2em}
\begin{algorithmic}\footnotesize
    \ALIRFor{$x \in A.x$}
        \ALIRFor{$y \in A.y \cap B.y$}
            \ALIRFor{$z \in B.z$}
                \State (x, y, z).
            \EndALIRFor
        \EndALIRFor
    \EndALIRFor
\end{algorithmic}
\vspace{-1em}
\caption{Inner join in ALIR.\label{fig:inner-join-alir}}
\end{wrapfigure}
Consider the following inner join and relational schema, which produces a tuple for every pair of tuples in $A$ and in $B$ that match on shared attributes: \[ R = A \ijoin B \quad\quad R(x, y, z); A(x, y); B(y, z) \] \Cref{fig:inner-join-alir} shows how the join can be represented in ALIR (for simplicity, duplicate handling is deferred until later). However, this is not the only way to encode this computation in ALIR. For example, the loops can be reordered arbitrarily without changing the semantics of the program, though depending on how the data is stored, not all orderings are valid; note that this is a limitation of the storage format, not our approach. In such cases, the loops must be rewritten in a compatible way. This restriction on data access orderings is captured by the coordinate tree.

\subsection{Loop specification}
\label{subsec:loop-specification}
The basic structure of ALIR consists of nested loops, each over a
specified domain. The domain of each loop is given as a multiset
expression, where the inputs come from levels of
the coordinate trees of the input relations.
The levels of a coordinate tree must be traversed in an order
compatible with how the data is physically laid out. For example,
consider a relation $A(x, y)$ sorted according to $(x, y)$ lexicographically.
To represent an ordered traversal, this relation can only be accessed as first $x$ then $y$,
that is, the values of $x$ are traversed in a loop nested outside
the loop that traverses $y$.

Our tool supports the following multiset operators, with standard
semantics~\cite{mathematics-of-multisets}: $\cup$ (union), $\cap$ (intersection), $-$ (multiset
difference), $+$ (multiset concatenation), $\times$ (Cartesian product)
as well as complement for inputs that are known to be
sets (have unique values).
Additionally, a special operator can be used to tag expressions
as needing additional inputs; this is notated as $A(B, C)$,
meaning $A$, but with additional inputs $B$ and $C$ (see \Cref{subsubsec:lookup-only-relations} for more details.)

\subsection{Virtual inputs}
\label{subsec:virtual-inputs}

In addition to input relations, ALIR also supports two types of virtual relations:
predicate expressions and singletons. Predicate expressions are virtual relations that only support lookup, not iteration, and are notated using set-builder notation.
This notation is used to represent the filter
operation. For example, the attribute $y$ of $\sigma_{y \text{ is even}}(B)$ has the domain
$y \in B.y \cap \{ y \mid y \text{ is even} \}$.
Singletons (notated $\{ v \}$) hold a single value $v$, and unlike predicate expressions,
support iteration.

\subsection{Precomputation and Temporary Relations}
\label{subsec:precomputation}

Within each loop (or outside all loops), ALIR also supports
precomputation, in which a temporary relation is created before the main computation is performed.
This can be useful for performance reasons, and it is also necessary for aggregation and non-equi joins
in general (temporary storage can be avoided in specific cases); precomputation is an explicit boundary for fusion.

For example, the expression $\sigma_{x^2 + y^2 = r^2}(A) \ijoin B$
that computes a filtered version of $A$ joined with $B$
could be computed by first
creating a temporary relation $T(x, y)$ corresponding to $\sigma_{x^2 + y^2 = r^2}(A)$, and then
performing an inner join between $T$ and $B$. This is shown in Figure~\ref{fig:alir-precompute}.
This can improve performance as we can skip all work in the inner loops
for values of $y$ where no matching $x$ exists.
The storage of temporary relations must also be specified in the
storage description.

\begin{figure}[b]%
    \begin{minipage}[t]{0.45\textwidth}
        \begin{algorithmic}\footnotesize
            \ALIRLet{$T(x, y)$}
                \ALIRFor{$x \in A.x$}
                    \ALIRFor{$y \in A.y \cap \{ y \mid x^2 + y^2 = r^2 \}$}
                        \State $(x, y).$
                    \EndALIRFor
                \EndALIRFor
            \ALIRIn
                \ALIRFor{$y \in T.y \cap B.y$}
                    \ALIRFor{$z \in B.z$}
                        \ALIRFor{$x \in T.x$}
                            \State $(x, y, z).$
                        \EndALIRFor
                    \EndALIRFor
                \EndALIRFor
            \EndALIRLet
        \end{algorithmic}
        \vspace{-1em}
        \caption{ALIR for $\sigma_{x^2 + y^2 = r^2}(A) \ijoin B$ where the left-hand side is precomputed into a temporary.}
        \label{fig:alir-precompute}
        \vspace{-0.5em}
    \end{minipage}\hfill\begin{minipage}[t]{0.45\textwidth}
        \begin{algorithmic}\footnotesize
            \ALIRLet{$T(\Sigma x, y)$}
                \ALIRFor{$x \in A.x$}
                    \ALIRFor{$y \in A.y \cap \{ y \mid x^2 + y^2 = r^2 \}$}
                        \State $T[y].\Sigma x \plusequals A.x$
                    \EndALIRFor
                \EndALIRFor
            \ALIRIn
                \ALIRFor{$y \in T.y \cap B.y$}
                    \ALIRFor{$z \in B.z$}
                        \ALIRFor{$\Sigma x \in T.\Sigma x$}
                            \State $(\Sigma x, y, z).$
                        \EndALIRFor
                    \EndALIRFor
                \EndALIRFor
            \EndALIRLet
        \end{algorithmic}
        \vspace{-1em}
        \caption{ALIR for $_{y}\agg_{\Sigma(x)}(A) \ijoin B$.}
        \label{fig:alir-aggregation}
        \vspace{-0.5em}
    \end{minipage}
\end{figure}

\subsection{Aggregation}
\label{subsec:aggregation}
Aggregate operations compute one or more aggregate functions (sum, product, etc.)\ across groups of values
that match on a set of grouping attributes.
To represent aggregation, the innermost expression is replaced with
a reduction operation. For example, the expression
$_{y}\agg_{\Sigma(x)}(A) \ijoin B$, which first aggregates
the sum of $x$ across the values of $y$ and then performs an inner join with $B$,
can be expressed in ALIR as shown in Figure~\ref{fig:alir-aggregation}.

This first performs the aggregation into a temporary relation $T$, then
performs the inner join to produce the final output. In general, aggregation is
always performed into a temporary, as intermediate values must be stored somewhere.
As such, for a fully-fused expression,
aggregation must occur at the outermost level.

\subsection{Outer joins}
\label{subsec:outer-joins}
\begin{figure}
    \begin{minipage}[t]{0.38\textwidth}
        \begin{algorithmic}\footnotesize
            \ALIRFor{$y \in A.y \cup B.y$}
                \ALIRFor{$x \in A.x \cup \varnothing$}
                    \ALIRFor{$z \in B.z \cup \varnothing$}
                        \State $(x, y, z).$
                    \EndALIRFor
                \EndALIRFor
            \EndALIRFor
        \end{algorithmic}
        \vspace{-1em}
        \caption{ALIR for the outer join $A \ojoin B$. $\cup \, \varnothing$ indicates
        loops that produce \NULL{} values.}
        \label{fig:alir-full-outer}
        \vspace{-1em}
    \end{minipage}
    \hfill
    \begin{minipage}[t]{0.58\textwidth}
        \begin{algorithmic}\footnotesize
            \ALIRFor{$y \in A.y\begingroup\color{red}(B.y)\endgroup$}
                \ALIRFor{$x \in A.x$}
                    \ALIRFor{$z \in B.z \cup \varnothing$}
                        \State $(x, y, z).$
                    \EndALIRFor
                \EndALIRFor
            \EndALIRFor
        \end{algorithmic}
        \vspace{-1em}
        \caption{ALIR for left join $A \ljoin B$. Without $B.y$ as an index-only relation (red), $B.z$ is uncorrelated with $A.y$ and therefore invalid.
        }
        \label{fig:alir-left-join-incorrect}
        \vspace{-1em}
    \end{minipage}
\end{figure}

Handling outer joins is essential to support the complete extended relational algebra
supported by most commercial RDBMSes. The behavior of outer joins is that the domain of the join
attribute is given by the left-hand operand (for left joins),
right-hand operand (for right joins), or the union of both (for full
outer joins). The remaining attributes are supplemented by \NULL{}s
where the join attribute is not present in a given relation.
The semantics of outer joins is shown in Figure~\ref{fig:joins}.

\hide{\centering
\begin{tabular}{cc}
    \multicolumn{2}{c}{$A$} \\ \toprule
    x & y \\ \midrule
    1 & 1 \\
    2 & 2 \\ \bottomrule
\end{tabular}\quad
\begin{tabular}{cc}
    \multicolumn{2}{c}{$B$} \\ \toprule
    y & z \\ \midrule
    2 & 3 \\
    3 & 4 \\ \bottomrule
\end{tabular}\par\vspace{1em}}

\begin{figure}
\begin{minipage}[b]{0.55\textwidth}
    \setlength{\tabcolsep}{0.3em}
    \footnotesize
    \adjustbox{valign=t}{\begin{tabular}{cc}\\
        \multicolumn{2}{c}{$A$} \\ \toprule
        x & y \\ \midrule
        1 & 1 \\
        2 & 2 \\ \bottomrule
    \end{tabular}}\quad
    \adjustbox{valign=t}{\begin{tabular}{cc}\\
        \multicolumn{2}{c}{$B$} \\ \toprule
        y & z \\ \midrule
        2 & 3 \\
        3 & 4 \\ \bottomrule
    \end{tabular}}\hfill
    \adjustbox{valign=t}{\begin{tabular}{ccc}
        \multicolumn{3}{c}{Inner} \\
        \multicolumn{3}{c}{$A \ijoin B$} \\ \toprule
        x & y & z \\ \midrule
        2 & 2 & 3 \\ \bottomrule
    \end{tabular}}\quad
    \adjustbox{valign=t}{\begin{tabular}{ccc}
        \multicolumn{3}{c}{Left} \\
        \multicolumn{3}{c}{$A \ljoin B$} \\ \toprule
        x & y & z \\ \midrule
        1 & 1 &   \\
        2 & 2 & 3 \\ \bottomrule
    \end{tabular}}\quad
    \adjustbox{valign=t}{\begin{tabular}{ccc}
        \multicolumn{3}{c}{Right} \\
        \multicolumn{3}{c}{$A \rjoin B$} \\ \toprule
        x & y & z \\ \midrule
        2 & 2 & 3 \\
          & 3 & 4 \\ \bottomrule
    \end{tabular}}\quad
    \adjustbox{valign=t}{\begin{tabular}{ccc}
        \multicolumn{3}{c}{Full} \\
        \multicolumn{3}{c}{$A \ojoin B$} \\ \toprule
        x & y & z \\ \midrule
        1 & 1 &   \\
        2 & 2 & 3 \\
          & 3 & 4 \\ \bottomrule
    \end{tabular}}
    \vspace{-1em}
    \caption{Semantics of different join types. \NULL{} is blank.}
    \label{fig:joins}
    \vspace{-1em}
\end{minipage}\hfill\begin{minipage}[b]{0.4\textwidth}
    \begin{algorithmic}\footnotesize
    \ALIRFor{$x \in A.x(C.x)$}
        \ALIRFor{$y \in (A.y \cap B.y)(C.y)$}
            \ALIRFor{$z \in B.z(C.z)$}
                \ALIRFor{$\dups \in (A \times B) - C$}
                    \State $(x, y, z).$
                \EndALIRFor
            \EndALIRFor
        \EndALIRFor
    \EndALIRFor
    \vspace{-1em}
    \caption{ALIR for $(A \ijoin B) - C$.}
    \label{fig:alir-dups}
    \vspace{-1em}
\end{algorithmic}
\end{minipage}
\end{figure}

To represent outer joins in ALIR, we only need to consider what the
requisite domain is for each nested loop. Non-join attributes are either
unchanged (both sides in inner join, LHS of left join, RHS of right join,
neither side in full outer join) or are possibly augmented with an additional \NULL{}
value. The join dimension (which combines both joined attributes) simply follows
the semantics of the corresponding join type. For example, a full outer join
$A \ojoin B$ can be expressed using the ALIR in Figure~\ref{fig:alir-full-outer}.
The special expression $\varnothing$ stores a single \NULL{} if the
other operand has no values.

\label{subsubsec:lookup-only-relations}
However, when computing left or right joins, we run into a problem:
consider the following ALIR shown in Figure~\ref{fig:alir-left-join-incorrect} for the left join $A \ljoin B$.
We would like the loop $z \in B.z \cup \varnothing$ to only iterate over
the values of $B.z$ for the tuples where $B.y = A.y$. However, without index-only relations, $B.y$ is not
even present in the IR! To solve this, we introduce the concept of
\emph{index-only relations}. These relations do not influence what values are being iterated over
and only introduce a branch in the
generated code depending on if the current value is present in that relation.

Note that this does not require the layer to have lookup capability:
it is also possible to perform co-iteration as usual, but the cases where
the current value is only present in the index-only relation are discarded.
As such, while the observed behavior (the values for which the
loop body is executed) is preserved, the actual iteration structure may be
altered by the presence of index-only relations.

\subsection{Duplicate handling}
\label{subsec:duplicate-handling}
Correctly handling duplicates is critical to ensuring correct results
for relational algebra, as extraneous or missing rows can cause
hard-to-diagnose errors, especially for aggregation. One particular example is the inner join of two single-attribute relations:
for example, if $A(x), B(y) = \lB 1, 1 \rB$, then
$A \ijoin_{x=y} B = \lB 1, 1, 1, 1 \rB$, according to the multiset
semantics of relational algebra, where $A \ijoin B \coloneqq
\sigma_{x=y}(A \times B)$ (cf.\ $A \cap B = \lB 1, 1 \rB$ as usual.)
The most flexible way to handle duplicates is to have a separate `duplicates' loop that generates the desired number
of duplicates. Then the distinction between intersection and join can be represented by having the duplicates loop perform either the intersection or Cartesian product of the duplicates.

For example, $A(x, y) \ijoin B(y, z) - C(x, y, z)$
cannot be represented using loops over only
multiset operations on $x, y, z$,
but can be represented in the ALIR in Figure~\ref{fig:alir-dups}.
Note that $C$ is an index-only relation
in every attribute, and the actual difference occurs in the
duplicate-handling loop.\footnote{There are other ways to perform this computation:
for example, one could make use of the fact that $A(x, y) - B(x, y)$
is equivalent to $\sigma_{x \notin B.x}(A) \cup \sigma_{y \notin B.y}(A)$ when $A, B$ are sets.}

\subsection{Non-equi joins}
\begin{figure}
    \begin{minipage}[b]{0.32\textwidth}
        \begin{algorithmic}\footnotesize
            \ALIRFor{$x \in A.x$}
                \ALIRFor{$y_1 \in A.y$}
                    \ALIRFor{$y_2 \in B.y \cap \{ y \mid y_1 < y \}$}
                        \ALIRFor{$z \in B.z$}
                            \ALIRFor{$\dups \in A \times B$}
                                \State $(x, y_1, y_2, z).$
                            \EndALIRFor
                        \EndALIRFor
                    \EndALIRFor
                \EndALIRFor
            \EndALIRFor
        \end{algorithmic}
        \vspace{-1em}
        \caption{\label{fig:alir-nonequi-inner}
            An inner non-equi join
            $A \ijoin_{A.y < B.y} B$.
        }
        \vspace{-1em}
    \end{minipage}
    \hfill
    \begin{minipage}[b]{0.32\textwidth}
        \begin{algorithmic}\footnotesize
            \ALIRFor{$x \in A.x(T.x)$}
                \ALIRFor{$y_1 \in A.y(T.y_1)$}
                    \ALIRLet{$T(y_2, z)$}
                        \State $\sigma_{A.y < B.y}(B)$
                    \ALIRIn
                        \ALIRFor{$y_2 \in T.y_2 \cup \varnothing$}
                            \ALIRFor{$z \in T.z \cup \varnothing$}
                                \ALIRFor{$\dups \in T \cup \varnothing$}
                                    \State $(x, y_1, y_2, z).$
                                \EndALIRFor
                            \EndALIRFor
                        \EndALIRFor
                    \EndALIRLet
                \EndALIRFor
            \EndALIRFor
        \end{algorithmic}
        \vspace{-1em}
        \caption{\label{fig:alir-nonequi-left}
            Left non-equi join for the same expression.
        }
        \vspace{-1em}
    \end{minipage}
    \hfill
    \begin{minipage}[b]{0.3\textwidth}
        \begin{algorithmic}\footnotesize
            \ALIRLet{$T(x, y_1, y_2, z)$}
                \State $A \ijoin_{A.y < B.y} B$
            \ALIRIn
                \State $A \ljoin_{A.y < B.y} B$ using $T$\dots
                \ALIRFor{$y_2 \in B.y(T.y_2)$}
                    \ALIRFor{$z \in B.z - T.z$}
                        \ALIRFor{$\dups \in B$}
                            \State $(\varnothing, \varnothing, y_2, z).$
                        \EndALIRFor
                    \EndALIRFor
                \EndALIRFor
            \EndALIRLet
        \end{algorithmic}
        \vspace{-1em}
        \caption{\label{fig:alir-nonequi-outer}
            Full outer non-equi join.
        }
        \vspace{-1em}
    \end{minipage}
\end{figure}
ALIR can also handle non-equi joins (joins over an arbitrary predicate, not just equality) in full generality.
An inner join $A \ijoin_\theta B$ with an arbitrary predicate $\theta$ is equivalent to
$\sigma_\theta(A \times B)$. Using this, an example non-equi join
$ A \ijoin_{A.y < B.y} B $
can be naturally expressed in ALIR, as shown in Figure~\ref{fig:alir-nonequi-inner}. There is one loop for each attribute, and the filter $A.y < B.y$ is inserted at the outermost valid location, which is the loop for $B.y$, resulting in the domain $B.y \cap \{ y \mid y_1 < y \}$.

However, a problem arises with \emph{non-inner} non-equi joins,
which cannot be described using a simple translation. The semantics of such joins is that
in addition to the tuples produced by the corresponding inner join,
tuples from the left relation (for left and full outer joins)
and/or right relation (for right and full outer joins)
are included if they would not otherwise be present,
with the missing values substituted with $\NULL$s.
The results of each type of join for the predicate $A.y < B.y$ on the data $A = \lB (1, 1), (3, 3) \rB, B = \lB (1, 2), (3, 4) \rB$ are shown below:

{\centering\footnotesize
\adjustbox{valign=t}{\begin{tabular}{cccc}
    \multicolumn{4}{c}{$A \ijoin_{A.y < B.y} B$} \\ \toprule
    x & A.y & B.y & z \\ \midrule
    1 &   1 &   3 & 4 \\ \bottomrule
\end{tabular}}\quad
\adjustbox{valign=t}{\begin{tabular}{cccc}
    \multicolumn{4}{c}{$A \ljoin_{A.y < B.y} B$} \\ \toprule
    x & A.y & B.y & z \\ \midrule
    1 &   1 &   3 & 4 \\
    3 &   3 &     &   \\ \bottomrule
\end{tabular}}\quad
\adjustbox{valign=t}{\begin{tabular}{cccc}
    \multicolumn{4}{c}{$A \rjoin_{A.y < B.y} B$} \\ \toprule
    x & A.y & B.y & z \\ \midrule
    1 &   1 &   3 & 4 \\
      &     &   1 & 2 \\ \bottomrule
\end{tabular}}\quad
\adjustbox{valign=t}{\begin{tabular}{cccc}
    \multicolumn{4}{c}{$A \ojoin_{A.y < B.y} B$} \\ \toprule
    x & A.y & B.y & z \\ \midrule
    1 &   1 &   3 & 4 \\
    3 &   3 &     &   \\
      &     &   1 & 2 \\ \bottomrule
\end{tabular}}\par\vspace{1em}}

In order to represent these operations,
we decompose the operation into an inner join that computes into a temporary
and using it to compute the full join, adding in the missing values.
Here, some form of temporary storage is unavoidable%
\footnote{Unless the data model is extended to allow for queries of the form
$\exists \upsilon \in B.\ \upsilon \mathrel{\theta} \tau.\alpha$
(i.e.,\ existence of a tuple where an attribute satisfies some
relation to another value.)}.
For example, we can compute the left join $A \ljoin_{A.y < B.y} B$
using the ALIR shown in Figure~\ref{fig:alir-nonequi-left}.
Note that this differs from a projected left join in duplicate handling:
a normal left join would produce the duplicate expression
$\dups \in T \times A \cup \varnothing$, which would produce extraneous
duplicates for every duplicate tuple in $A$.

The same idea can be used for full outer joins, but when augmenting
the result with $B$, we need to iterate over
$A.x(T.y) \cup \varnothing$ in the outermost loop. As the presence of
$\varnothing$ depends on values only produced in a later loop, this cannot
be used to generate code. To fix this, we append an extra loop with iteration
order switched, which avoids this problem. This is shown in Figure~\ref{fig:alir-nonequi-outer}.

\section{Iteration Machines}
\label{sec:iteration-machines}

To generate loops for single-dimensional iteration, we convert
multiset expressions to \emph{iteration machines} to represent the
structure of the loops.
Iteration machines are inspired by iteration lattices as used in TACO~\cite{kjolstad2017,henry2017}, but there are several key differences. First, we have created a new, simpler formulation using deterministic finite automata (DFAs), a new construction algorithm that is generalized to handle a larger class of expressions, and an optimization algorithm that is guaranteed to minimize the complexity of the generated loops. Second, iteration machines are able to generate correct code for multiset operations, particularly the multiset difference and Cartesian product. Finally, iteration machines support operations necessary for generation of outer joins and non-equi joins.

Iteration machines are a data structure that concisely represents exactly the information necessary to generate loops for co-iteration of a given multiset expression. It shows how the iteration evolves as the inputs run out of values by storing a set of \emph{nodes} at which the iteration structure changes. Additionally, it contains the value of the expression at these nodes, which is then used to determine how to progress iteration in the body of the generated loops. The use of a single data structure for both cases simplifies reasoning about the iteration structure, and in both cases the same minimized iteration machine is optimal in terms of generated code complexity.

To give intuition for how iteration machines are used to generate
loops for co-iteration,
we first look at a simple example of $(A \cap B) \cup C$, which results in an iteration machine that is also an iteration lattice.
Iteration begins by simultaneously co-iterating all input multisets.
Generating the code inside the loop to handle the cases depending on
the presence of each input is covered in the next section; we first
look at the structure of the loops themselves.
To avoid performing unnecessary work, as soon as an input multiset
runs out of values, we would like to remove it from the loop.
Consider the case where $A$ (or symmetrically, $B$) runs out of
values. Then we only need to resume iterating over $C$ by itself,
as it does not matter what $B$ contains. However, if $C$ runs out of
values, we must iterate over the remaining elements of $A \cap B$,
which requires iterating over both $A$ and $B$. Finally, if both
$C$ and either of $A$ or $B$ run out, we should stop iteration.

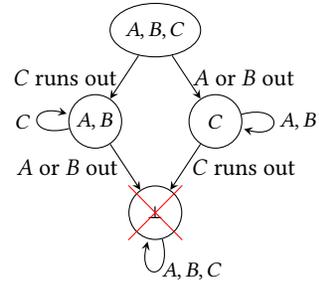
\begin{wrapfigure}[12]{r}{0.4\textwidth}\centering\vspace{-1.2em}\footnotesize
\begin{tikzpicture}[y=1.5cm,scale=0.8]
        \node[ellipse,draw,minimum size=2.5em, inner sep=1pt] (n0) at (0, 0) {$A, B, C$};
        \node[circle,draw,minimum size=2.5em, inner sep=1pt] (n1) at (-1, -1) {$A, B$};
        \node[circle,draw,minimum size=2.5em, inner sep=1pt] (n2) at (1, -1) {$C$};
        \node[ellipse,draw,minimum size=2.5em, inner sep=1pt] (n3) at (0, -2) {$\bot$};
            \node[draw=red,cross out,minimum size=2.5em] at (n3) {};

        \draw[->,>=stealth] (n0) -- node[left] {\small$C$ runs out} (n1) (n1) edge node[left] {\small$A$ or $B$ out} (n3);
        \draw[->,>=stealth] (n0) -- node[right] {\small$A$ or $B$ out} (n2) (n2) edge node[right] {\small$C$ runs out} (n3);
        \draw[->,>=stealth,loop left] (n1) edge node[left] {$C$} ();
        \draw[->,>=stealth,loop right] (n2) edge node[right] {$A, B$} ();
        \draw[->,>=stealth,loop below] (n3) edge node[right] {$A, B, C$} ();
    \end{tikzpicture}\vspace{-1.5em}
    \caption{Iteration machine for $(A \cap B) \cup C$. $\bot$ represents the end of iteration. Self-edges are added for completeness as a DFA.}
    \label{fig:exmach}
\end{wrapfigure}
The structure of how the iteration changes in this example is shown in the
diagram in Figure~\ref{fig:exmach}. The diagram is an example of an \emph{iteration machine}. An iteration
machine precisely specifies how the set of inputs to iterate over
changes as inputs run out of values. It is represented as a
deterministic finite automaton whose alphabet is the set of all
inputs.
The states of the DFA are labeled with what inputs must be
co-iterated over, and the transition function describes how to move
between states as each input runs out.

Iteration machines must have a particular structure to ensure
it is possible to iterate in a defined order without repeats.
In particular, iteration machines must satisfy the following properties:

\begin{enumerate}
    \item States must be unique. No two states can iterate over the exact same multisets.
    \item The initial state must be the set of all inputs.
    \item All transitions from a state $S$ across input $x \notin S$
        must be a self-loop.
    \item For a transition from $S_1$ to $S_2$ across input $x$,
        $S_2 \subseteq S_1 - \{x\}$. That is, inputs cannot be added,
        and the set that runs out must be removed. Also, there must
        not be an intermediate state $S'$ between $S_1$ and $S_2$:
        $\nexists S'.\, S_2 \subset S' \subset S_1$. \label{item:im-transition}
    \item For every two states $S_1$ and $S_2$, the state $S_1 \cup S_2$
        must also exist. This is necessary for case handling to be
        defined on this set of inputs.
\end{enumerate}
These properties give iteration machines a (bounded) join-semilattice
structure, which is used to determine how to separate cases within the loop.

\subsection{Case Handling}
Within a loop, the compiler must generate different code, splitting each iteration into multiple cases depending on what relations contain the current value. This process of \emph{case handling} involves determining what cases need to be handled, the conditions for each case, and how iteration proceeds for each case.
The same iteration machine used to generate loop structure can be used
as-is for case handling: for any possible set of inputs that contain
a given value, an iteration machine assigns a representative state
(node).
Indeed, this is the key reason why iteration machines are useful for
code generation: iteration machines precisely capture the behavior of
an expression by splitting the space of combinations of input relations (Venn diagram) into disjoint spaces each
represented by a state. Then the progression of iteration as inputs run out
can be derived by looking at what subspace remains reachable as we move
down the machine.

For example, in the example of $A \cap B \cup C$, the set of inputs
$\{A, C\}$ can be handled exactly the same as the set $\{C\}$, as once
$B$ runs out of values, it does not matter whether a value is in $A$
or not. Therefore, we assign both of these sets the representative of
$\{C\}$. And by looking at the iteration machine, it is clear that this
case only needs to be checked in states $(A, B, C)$ and $(C)$.

For an iteration machine and set of inputs,
the representative is given by the \emph{floor function}.

\begin{restatable}[Floor function]{lemma}{lemfloor}\label{lem:floor}
    Given an iteration machine over input set $S$, for every combination of inputs
    $p \subseteq S$, there exists a unique floor node with label $\floor(p)$
    such that $\floor(p) \subseteq p$ and $\nexists n'.\,
    \floor(p) \subseteq n' \subseteq p$. Additionally, $\floor$
    is monotone.
\end{restatable}

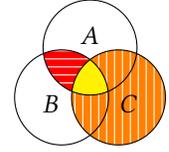
\begin{wrapfigure}{r}{0.2\textwidth}\centering
    \vspace{-2em}
    \begin{tikzpicture}[scale=0.5]
        \begin{scope}
            \clip \vennThreeFirst;
            \fill[preaction={fill=red},pattern=horizontal lines,pattern color=white] \vennThreeSecond;
        \end{scope}
        \begin{scope}
            \fill[preaction={fill=orange},pattern=vertical lines,pattern color=white] \vennThreeThird;
        \end{scope}
        \begin{scope}
            \clip \vennThreeFirst;
            \clip \vennThreeSecond;
            \fill[yellow] \vennThreeThird;
        \end{scope}
        \draw \vennThreeFirst;
        \draw \vennThreeSecond;
        \draw \vennThreeThird;
        \node at (90:1.2cm) {$A$};
        \node at (210:1.2cm) {$B$};
        \node at (330:1.2cm) {$C$};
    \end{tikzpicture}
    \caption{Venn diagram for the expression $A \cap B \cup C$;
    sections colored according to IM nodes.
    }\vspace{-3em}
    \label{fig:venn}
\end{wrapfigure}

Note that due to uniqueness,
the transition function for an iteration machine
is uniquely determined from the set of states.
In particular,
    $\delta(n, x) = \floor(n - \{x\})$.
This follows from requirement \ref{item:im-transition} and
uniqueness of the floor function.
This is further extended by the following lemma, which ties together
the transition function and the floor function:
\begin{restatable}[Coherence of transition function]{lemma}{lemlatticedfa}\label{lem:latticedfa}
    For an iteration machine $M$ over $S$, the state after
    a sequence of transitions $a_1, \dots, a_n$ is
    $\floor(S - \{ a_i \mid 1 \le i \le n \})$.
\end{restatable}

To give intuition for what this function represents,
we look at the Venn diagram in Figure~\ref{fig:venn}.
We color in this Venn diagram according to the (non-empty) nodes
of the given iteration machine for $A \cap B \cup C$, bottom up.
We first color $C$ orange, then $A \cap B$ (representing $(A, B)$)
red. Finally, $A \cap B \cap C$ is colored yellow. These colors
represent which node is a representative at any given point in the
Venn diagram: even though $A \cap C$ is not present in the IM,
we see that it is covered by its representative $C$.

Each node in an iteration machine may be labeled with one of three
possible behaviors: producer (any input with this representative
should produce a value), omitter (this input should not produce a value),
or not ready (the machine is not immediately ready to produce output,
but there may still be values remaining.) The not ready state is
discussed in more detail in the next section, as it is particular
to multiset semantics. Let $\interp(n)$ represent the label of the node $n$,
where $\top$ represents producer, $\bot$ omitter, and $\emptyset$
not ready.

Using this intuition, the floor function gives a natural
interpretation of an iteration machine:
    $\lBrack M \rBrack(p) = \interp(\floor(p)).$
Two iteration machines have identical behavior if they have the same
interpretation. See \Cref{subsec:optimization} for an example of two distinct machines with the same behavior.

\subsection{Multiset Semantics}\label{sec:multiset}
\begin{figure}
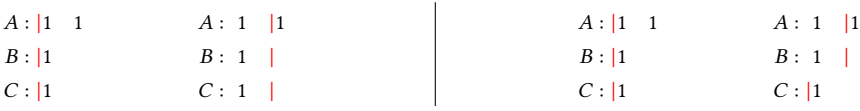

    \begin{minipage}[t]{0.45\textwidth}\footnotesize
        \setlength{\abovedisplayskip}{0pt}
        \setlength{\belowdisplayskip}{0pt}
        \setlength{\abovedisplayshortskip}{0pt}
        \setlength{\belowdisplayshortskip}{0pt}
        \begin{align*}
            A&: \textcolor{red}{|}1          \quad 1 &\quad A&: \hphantom{|}1          \quad \textcolor{red}{|}1 \\
            B&: \textcolor{red}{|}1\hphantom{\quad 1}&\quad B&: \hphantom{|}1\hphantom{\quad}\textcolor{red}{|}\hphantom{1} \\
            C&: \textcolor{red}{|}1\hphantom{\quad 1}&\quad C&: \hphantom{|}1\hphantom{\quad}\textcolor{red}{|}\hphantom{1}
        \end{align*}
    \end{minipage}
    \hfill\vline\hfill
    \begin{minipage}[t]{0.45\textwidth}\footnotesize
        \setlength{\abovedisplayskip}{0pt}
        \setlength{\belowdisplayskip}{0pt}
        \setlength{\abovedisplayshortskip}{0pt}
        \setlength{\belowdisplayshortskip}{0pt}
        \begin{align*}\footnotesize
            A&: \textcolor{red}{|}1          \quad 1 &\quad A&: \hphantom{|}1          \quad \textcolor{red}{|}1 \\
            B&: \textcolor{red}{|}1\hphantom{\quad 1}&\quad B&: \hphantom{|}1\hphantom{\quad}\textcolor{red}{|}\hphantom{1} \\
            C&: \textcolor{red}{|}1\hphantom{\quad 1}&\quad C&: \textcolor{red}{|}1\hphantom{\quad}\hphantom{|}\hphantom{1}
        \end{align*}
    \end{minipage}
    \caption{Diagram showing iteration state for (left) incorrect iteration machine (right) correct iteration machine. Red bars indicate the current position of each iterator, where the element after the bar is the current value. The state on the left is the initial state, and the state on the right shows the state after a single iteration.}
    \label{fig:multiset-iterators}
\end{figure}

Applying this procedure directly to multisets works well for
multiset union ($\#(A \cup B)(x) = \max(\#A(x), \#B(x))$,
where $\#M(x)$ means the number of times $x$ appears in the multiset $M$) and
intersection ($\#(A \cap B)(x) = \min(\#A(x), \#B(x))$), but a problem
arises when we try to handle difference ($\#(A - B)(x) = \max(0,
\#A(x) - \#B(x))$).
To illustrate this problem, we look at the expression $(A - B) \cap C$
on the inputs $A = \lB 1, 1 \rB, B = \lB 1 \rB, C = \lB 1 \rB$.
The correct result should be $(A - B) \cap C = \lB 1 \rB$.
A minimal iteration machine with interpretation
\[
    \lBrack M \rBrack(p) = \begin{cases}
        \top & \text{if } x \in (A - B) \cap C \\
        \bot & \text{otherwise}
    \end{cases}
\] (for arbitrary $x$
whose membership in $A, B, C$ has representative $p$)
has nodes $\{ (A, B, C), (A, C), () \}$ and interpretation
$\interp(A, B, C) = \bot, \interp(A, C) = \top, \interp() = \bot$.

But applying this iteration machine produces the incorrect behavior shown in Figure~\ref{fig:multiset-iterators} (left).
First, since 1 appears in all three inputs, it falls into
representative $(A, B, C)$, which has interpretation $\bot$.
To proceed, all three inputs are incremented, since they all contain
1. However, the next state has representative $()$, since a second 1
is only contained in $A$; since $\interp() = \bot$,
the output is $\lB \rB$.

This problem arises due to misalignment resulting from empty spaces
from the difference operator.
To fix this, we can treat these empty spaces separately, and instead
fast-forward until a value is ready or we run out of values;
equivalently, we delay processing the inputs not involved in the
difference until we are past the empty space.
This is the purpose of the ``not ready'' behavior. Additionally,
we must indicate that only inputs involved in the difference should
be incremented. This correct behavior is shown in Figure~\ref{fig:multiset-iterators} (right).

In the correct IM, $(A, B, C)$ is instead labeled with behavior
$\emptyset(A, B)$, meaning that only $A$ and $B$ should be incremented.
As a result, in the second state, the resulting representative is
$(A, C)$, which produces the value 1, as desired.

\emph{Disjoint union.}
In many cases, disjoint union is not combined with other operators and the iteration order does not matter; in such cases, it is best to use two separate loops: for example, a loop over $A + B$ is equivalent to a loop over $A$ followed by a loop over $B$. However, when disjoint union is combined with other operators, or when the iteration order must be preserved, the disjoint union operator must be handled in the iteration machine. It is treated similarly to union, except that the behavior of nodes where both sides are present should only advance one side, leaving the other side to be iterated in the next iteration. This requires an additional behavior, $\top(\vec{N})$, with the $\vec{N}$ parameter analogous to $\emptyset(\vec{N})$.

\emph{Multiple cursors.}
Multiple cursors are necessary in certain situations, where the same input needs to be both advanced and not advanced.
For example, consider the expression
$(A - B) \cup (C - A)$.
In the $(A, C)$ state, the right-hand side of the union is not ready, so we only want to advance the
inputs on the right-hand side ($A$ and $C$.) However, $A$ is also present on the left-hand side.
To solve this, we must split iteration over $A$ into multiple cursors. Then we only advance the cursor
corresponding to $A$ as it appears in the right-hand side.

\subsection{Generation}
Iteration machines can be generated bottom-up, from single
set to expressions, using the product construction for DFAs.
In particular, the product construction simplifies generation
compared to iteration lattices, as getting the correct behavior does not require any special case handling.
Iteration machines capture all essential properties of the desired loops by construction.

\begin{description}

\item[Segment rule]
The iteration machine for a single set $A$ has two nodes:
$(A)$ and $()$, where $\interp(A) = \top(A), \interp() = \bot$. (See \Cref{sec:multiset} for a description of the notation $\top(\vec{N})$.)

\item[Complement rule]
The complement of an iteration machine can obtained by simply
reversing the behavior of all nodes ($\top$ to $\bot$ and vice versa.)
Since complement is not well-defined for multisets, we do not need to
handle $\emptyset$.

\item[Product construction]
To generate the iteration machine for a binary operator
$E_1 \circ E_2$, we use a product construction similar to that for
DFA intersection and union described below.
\end{description}
\hide{
\begin{minipage}{0.1\textwidth}\centering
        \begin{tikzpicture}[y=1.5cm,scale=0.7,every node/.style={font=\footnotesize}]
            \node[circle,draw,minimum size=2.5em, inner sep=1pt] (n2) at (0, 0) {$A$};
            \node[circle,draw,minimum size=2.5em, inner sep=1pt] (n3) at (0, -1) {$\bot$};
                \node[draw=red,cross out,minimum size=2.5em] at (n3) {};
    
            \draw (n2) -- (n3);
        \end{tikzpicture}
        \captionof{figure}{Segment rule for $A$.}
        \label{fig:segrule}
\end{minipage}
}

Depending on the operator $\circ$, the interpretation for a resulting node $u_1 \cup u_2$ in the product construction
is given by $\interp_1(u_1) \circ \interp_2(u_2)$, defined as follows:
\begin{itemize}
    \item $\emptyset(\vec{N}) \circ \emptyset(\vec{M}) = \emptyset(\vec{N} \cup \vec{M})$.
    \item $\emptyset(\vec{N}) \circ v = v \circ \emptyset(\vec{N}) = \emptyset(\vec{N})$,
        for $v \in \{\top(\vec{M}), \bot\}$, except $\emptyset(\vec{N}) \cap \bot = \bot$.
    \item$\top(\vec{N}) + \top(\vec{M}) = \top(\vec{N})$. Otherwise, $+$ behaves the same as $\cup$.
    \item $\top(\vec{N}) - \top(\vec{M}) = \emptyset(\vec{N} \cup \vec{M})$. In set semantics,
        $\top - \top = \bot$, so that complement is well-defined.
    \item$\top(\vec{N}) \cap \top(\vec{M}) = \top(\vec{N}) \cup \top(\vec{M}) = \top(\vec{N} \cup \vec{M})$.
    \item $\top(\vec{N}) \cup \bot = \bot \cup \top(\vec{N}) = \top(\vec{N}) - \bot = \top(\vec{N})$.
    \item Otherwise, $\interp_1 \circ \interp_2$ is given by the standard interpretation of
        $\cup, \cap$, $-$, etc.
\end{itemize}
The Cartesian product $\times$ behaves the same as intersection, but instead of $u_1 \cup u_2$, the node is instead labeled with $\{ R_1 \times R_2 \mid R_1 \in u_1, R_2 \in u_2 \}.$ These nodes require special handling in iteration.

The full procedure for product construction is given in Algorithm~\ref{alg:product}.
The algorithm performs a breadth-first search on the product DFA by simultaneously traversing the left and right iteration machines,
where $Q$ is the queue of pairs to visit. For each pair $u_1, u_2$, the resulting node is labeled with the union $u_1 \cup u_2$ (if not product), and its interpretation is given as described above. Then, we proceed on each transition $r$ in $u_1 \cup u_2$, where the result in both IMs is given by Lemma~\ref{lem:latticedfa}. Finally, once all nodes are discovered, the produced iteration machine is returned.

Note that we do not lose information by combining $(u_1, u_2)$ as $u_1 \cup u_2$.
This is shown in the following lemma, which guarantees that there is at most one reachable node
with each label $u_1 \cup u_2$, so no distinct reachable nodes are merged together:
\begin{restatable}[Uniqueness of nodes in product construction]{lemma}{lemdfauniq}\label{lem:dfauniq}
    If $u_1, v_1 \in N_1$ and $u_2, v_2 \in N_2$ such that
    $u_1 \cup u_2 = v_1 \cup v_2$, either $(u_1, u_2) = (v_1, v_2)$
    or one or both of $(u_1, u_2), (v_1, v_2)$ are not reachable
    in the product DFA.
\end{restatable}

\begin{algorithm}[tbp]
    \caption{Product construction for IM generation}
    \label{alg:product}
    \begin{algorithmic}\footnotesize
        \Procedure{Product}{$N_1, \interp_1, N_2, \interp_2, \circ$}
            \State $Q \gets \{ (\top_1, \top_2) \}$ \Comment{$\top$ is the maximum element}
            \State $N, \interp, E \gets \{\}, \{\}, \{\}$ \Comment{edge info is used for minimization}
            \While{$Q$ is not empty}
                \State $u_1, u_2 \gets \operatorname{pop}(Q)$
                \State $l \gets u_1 \cup u_2$, or $\{ R_1 \times R_2 \mid R_1 \in u_1, R_2 \in u_2 \}$ if $\circ$ is $\times$
                \State skip to next pair if $l \in N$; otherwise add $l$ to $N$
                \State $\interp(l) \gets \interp_1(u_1) \circ \interp_2(u_2)$
                \For{$r \in l$} \Comment{where an input $r \in l$ if $r$ is contained in any product}
                    \State $v_1, v_2 \gets \floor_1(u_1 - \{r\}), \floor_2(u_2 - \{r\})$
                    \State add edge $(u_1, u_2) \to (v_1, v_2)$ to E
                    \State $Q \gets Q \cup \{(v_1, v_2)\}$
                \EndFor
            \EndWhile
            \State \Return $N, \interp, E$
        \EndProcedure
    \end{algorithmic}
\end{algorithm}

\subsection{Optimization}
\label{subsec:optimization}
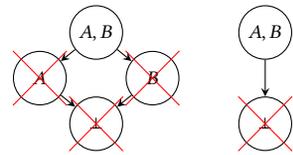
\begin{wrapfigure}[7]{r}{0.4\textwidth}\centering\vspace{-3em}
    \begin{tikzpicture}[x=1.5cm, y=1.2cm,scale=0.5,every node/.style={font=\scriptsize}]
        \scope[shift={(-1.5,0)}]
            \node[circle,draw,minimum size=2em, inner sep=1pt] (n1) at (0, 0) {$A, B$};
            \node[circle,draw,minimum size=2em, inner sep=1pt] (n2) at (-1, -1) {$A$};
                \node[draw=red,cross out,minimum size=2em] at (n2) {};
            \node[circle,draw,minimum size=2em, inner sep=1pt] (n4) at (1, -1) {$B$};
                \node[draw=red,cross out,minimum size=2em] at (n4) {};
            \node[circle,draw,minimum size=2em, inner sep=1pt] (n3) at (0, -2) {$\bot$};
                \node[draw=red,cross out,minimum size=2em] at (n3) {};

            \draw[->,>=stealth] (n1) -- (n2); \draw[->,>=stealth] (n2) -- (n3);
            \draw[->,>=stealth] (n1) -- (n4); \draw[->,>=stealth] (n4) -- (n3);
        \endscope

        \scope[shift={(1.5,0)}]
            \node[circle,draw,minimum size=2em, inner sep=1pt] (n1) at (0, 0) {$A, B$};
            \node[circle,draw,minimum size=2em, inner sep=1pt] (n3) at (0, -2) {$\bot$};
                \node[draw=red,cross out,minimum size=2em] at (n3) {};

            \draw[->,>=stealth] (n1) -- (n3);
        \endscope
    \end{tikzpicture}\vspace{-0.5em}
    \caption{Two distinct IMs for $A \cap B$, with omitter nodes crossed out. Edge labels and self-loops are
    omitted.}
    \label{fig:minex}
\end{wrapfigure}

Consider the two iteration machines (IMs) in Figure~\ref{fig:minex}. Both IMs represent the same
expression $A \cap B$, but the second has more nodes than the first.
To speed up iteration and reduce the amount of generated code, it is desirable to ensure that only a minimal
number of nodes is generated. To do this, we run a minimization step
after product construction.

Consider a node $n$. It can be removed only if removing it produces
a valid IM $M'$, and for all $p$,
\[ \floor(p) = n \implies (\interp'(\floor'(p)) = \interp(\floor(p))), \]
where $\floor'(p)$ is the node corresponding to $p$ in $M'$.
In fact, this is true whenever $\interp(n) = \interp'(\floor'(n))$.
Observe also that removing a node can never cause another node below
it (i.e.\ a subset) to be removable, so removability only flows upwards.
Therefore, we can minimize iteration machines by traversing from
bottom to top, as shown in Algorithm~\ref{alg:minim}.

\begin{algorithm}[tbp]
    \caption{Iteration machine minimization}
    \label{alg:minim}
    \begin{algorithmic}\footnotesize\Procedure{Minimize}{$N, \interp, E$}
        \State $Q \gets \text{nodes with zero out-degree}$
        \State $\operatorname{ip}(n) \gets \{\} \text{ for all $n \in N$}$ \Comment{holds the immediate predecessor (equiv. $\floor'(n)$) for each node}
        \While{$Q$ is not empty} \Comment{topological sort on reverse graph}
            \State $v \gets \operatorname{pop}(Q)$
            \If{$\operatorname{ip}(v) \ne v$ and $\interp(\operatorname{ip}(v)) = \interp(v)$}
                \State remove $v$ from machine
            \EndIf
            \ForAll{$(u, v) \in E$}
                \State reduce $\operatorname{deg}(u)$ by 1 and add to $Q$ if zero
                \State $\operatorname{ip}(u) \gets \operatorname{ip}(u) \cup v$
            \EndFor
        \EndWhile
    \EndProcedure\end{algorithmic}
\end{algorithm}

\subsection{Index-only relations}
During construction, nodes in the IM are tagged with associated index-only relations: to construct the IM for $E_1(E_2)$, we tag each node (except $\bot$) in the IM for $E_1$ with all relations in $E_2$, and this is propagated in product construction: a node $u_1 \cup u_2$ is tagged with the relations for $u_1$ and for $u_2$. After optimization, nodes are re-introduced: for each node $u_1(\vec{N})$, we construct the nodes $u_1 \cup n$ for $n \subseteq \vec{N}$; the interpretation is given by $\interp(\floor(u_1 \cup n))$. This ensures that every combination of $\vec{N}$ has a distinct state and thus branch in code generation.

\subsection{Legality}
\label{sec:legality}

There are two cases where a multiset expression cannot be iterated over. First, consider the expression $\bar{A}$; this requires iterating over every value \emph{not} in $A$. This case occurs when the final node ($\bot$) is a producer. In order for the expression to be well-defined, there must be a universal set that contains all possible values. In some cases, iteration over the universal set is possible, such as a finite range of integers, but for most data types, the universal set is not iterable (e.g.\ all strings).

Second, consider iterating over a relation $A$ whose storage does not have the iterator capability, only lookup. Since iteration requires some node to have the iterator capability, it is not possible to generate code in this case. Either the storage for $A$ must be changed to one with iterator capability, or the iteration machine must be modified as described below. This case occurs when any node in the IM, except $\bot$, has no relations with the iterator capability.

In either case, we describe such iteration machines as \emph{illegal}, and they must be \emph{legalized} to produce a legal iteration machine in order to generate code. To legalize an iteration machine, each node is augmented with the universal set $\mathbb{U}$. Then a new final node is added (with omitter behavior, but this is never reachable). The legalized iteration machine then can be used to generate code which iterates through the universal set. If it is not possible to iterate over the universal set, then iteration for the expression is not possible.

\section{Code Generation}
\label{sec:code-generation}

In order to generate executable code from ALIR, two additional pieces are necessary as presented above.
First, we need to generate an iteration machine for each loop domain, as this tells us the structure of the generated loops. Second, we need information on how to generate code for each layer, according to the interface in our data model. Given this information, we can then proceed to generate optimized C++ code.

\begin{algorithm}[tbp]
    \caption{Code generation for loops}
    \label{alg:codegen}
    \begin{algorithmic}\footnotesize
        \Procedure{GenerateLoop}{loop domain $D$, body, storage description for each layer}
            \State $M \gets $ generate iteration machine for $D$
            \State emit $\langle R.\texttt{init}() \rangle$ for each layer $R$
            \For{each non-empty state $S$ in $M$ in topological order}
                \State emit \textbf{while} ($\bigwedge_{R \in \textsf{iterators} \cap S} \langle R.\texttt{valid}() \rangle$) \{
                \State compute minimum value $m$ of $\langle R.\texttt{curval}() \rangle$ for each iterator layer
                \State compute galloping value $g$ (Subsection~\ref{subsec:galloping})
                \If{$g \ne -\infty$}
                    \State emit \textbf{if} ($m \ne g$) \{ advance all iterators to $g$ using \texttt{skipto}; \textbf{continue} \}
                \EndIf
                \For{each reachable state $T$ from $S$ in topological order}
                    \State emit \textbf{if} ($\bigwedge_{R \in T} \langle R.\texttt{present}() \rangle$) \{
                    \If{$\interp(T) = \top$}
                        \State emit body for state $T$
                        \State advance iterators in $T$
                    \Else \Comment{$\interp(T) = \bot, \emptyset$}
                        \State advance appropriate iterators (either all iterators in $T$ or $\vec{R}'$ if $\interp(T) = \emptyset(\vec{R}')$.
                    \EndIf
                    \State emit \}
                \EndFor
                \State emit \}
            \EndFor
        \EndProcedure
    \end{algorithmic}
\end{algorithm}

\subsection{Loop generation}

The most important part of the code generation algorithm is loops. The code generation algorithm for loops is shown in Algorithm~\ref{alg:codegen} and the generated code structure is as follows:

\begin{algorithmic}\footnotesize
    \State for each input relation $R$ that appears in the loop domain, $\langle R.\texttt{init}() \rangle$
    \While{$\bigwedge_{R \in \textsf{iterators}} \langle R.\texttt{valid}() \rangle$}
            \Comment{loop corresponding to root (topmost) state, containing all inputs}
        \State $\textit{min} \gets \min_{R \in \textsf{iterators}}(\langle R.\texttt{curval}() \rangle)$
        \State based on current values, skip past values that cannot produce values (\ref{subsec:galloping})
        \State handle cases and advance iterators (\ref{subsec:case-handling})
    \EndWhile
    \While{$\bigwedge_{R \in S_2 \cap \textsf{iterators}} \langle R.\texttt{valid}() \rangle$}
            \Comment{loop corresponding to state $S_2$}
        \State $\textit{min} \gets \min_{R \in S_2 \cap \textsf{iterators}}(\langle R.\texttt{curval}() \rangle)$
        \State \dots
    \EndWhile
    \State \dots{} loops for remaining states \dots
\end{algorithmic}

After generating the iteration machine, the algorithm initializes variables as necessary for each relation. Then, for every state in the iteration machine in topological order, a loop is generated with the condition corresponding to the required iterators for that state. During iteration, the current value being processed is indicated by the minimum value (\textit{min}) of all iterators. Next, galloping (\Cref{subsec:galloping}) is performed on relevant iterators to skip past values when possible. Finally, the algorithm emits a series of branches to determine the correct behavior based on which relations contain \textit{min}. Within each branch, the loop body is emitted if the corresponding state is a producer, and the appropriate iterators are advanced.

As described in \Cref{sec:iteration-machines}, we generate multiple concrete loops for a given abstract loop, to save computation as relations run out of values. Each state in the iteration machine represents one loop to be generated, with the exception of states where no producer node is reachable.

By ordering the loops in topological order, control flow moves automatically between loops, avoiding the need for complex control flow based on which relations are valid.
\begin{restatable}{lemma}{lemctrlflow}
    If loops are arranged in topological order, after exiting a loop, the next loop whose condition is met corresponds to the correct state in the iteration machine.
\end{restatable}
In fact, ordering the loops by the number of remaining inputs is sufficient. By using this ordering, we can use the sequencing of loops to automatically move between loops according to the IM.

\subsection{Case handling}
\label{subsec:case-handling}

Case handling is done in a similar way to loop generation. We generate
branches for each case according to the iteration machine.
For each branch, we test if the current iteration state matches the
given state. Since the current value being iterated over is
\textit{min} as computed at the beginning of the loop body, we test
each input relation $R$ using the method
$R.\texttt{present}(\textit{min})$.
We know that we are in a branch
if all relations in its label are present, and the current iteration
state does not already fall into a more specific state.
To handle the overlapping of states, it is sufficient to test each
state from most specific to least specific, relying on fallthrough to
prevent a less specific state from shadowing a more specific state.

Additionally, we only need to handle those which are reachable from
the state corresponding to the current loop, as any other state would
involve an input which has already run out. Due to the structure of
iteration machines, any state reachable from a given other state must
have a label which is a subset of that state (by
Lemma~\ref{lem:latticedfa}). This lets us enumerate the necessary cases to handle.

At the end of each case, we need to advance the appropriate iterators. For $\bot$ case, this is every iterator in the corresponding state, and for $\top(\vec{R})$ and $\emptyset(\vec{R})$, this is $\vec{R}$. Product relations (as produced by the product construction for $\times$) require special handling, as shown in Algorithm~\ref{alg:prodadv}.

\begin{algorithm}[tbp]
    \caption{Advance operation for product relations}
    \begin{algorithmic}\footnotesize
        \Procedure{AdvanceProduct}{$R_1, \dots, R_n$}
            \State emit $\langle R_n.\texttt{advance}() \rangle$
            \For{$i = n-1, \dots, 1$}
                \State emit \textbf{if} ($!\langle R_{i+1}.\texttt{valid}() \rangle$) \{
                $\langle R_{i+1}.\texttt{reset}() \rangle$
                ; $\langle R_i.\texttt{advance}() \rangle$
                \}
            \EndFor
        \EndProcedure
    \end{algorithmic}
    \label{alg:prodadv}
\end{algorithm}

\subsection{Body code}

Within each branch, we perform different actions depending on the
behavior of the state associated with that branch. If the state is an
omitter, we simply advance the iteration state and proceed to the next
iteration. Similarly, if the state is not ready, we only advance the
iterators involved in that state. Finally, if the state is a producer,
we generate code for the loop body.

There are three types of operations in a loop body: nested
loops, producing an output tuple, or aggregation.
For nested loops, we simply proceed recursively.
Producing an output tuple is similarly straightforward: we populate
a result tuple with the values specified and append it to the output.

Aggregation is slightly more complex. We first populate a key tuple
with the attributes used for grouping. This key tuple is used to look
up the corresponding value tuple for aggregation. If it is not found,
a new tuple is inserted as the identity of the aggregation operations.
Finally, the looked up or inserted tuple is updated using the values
specified. As an optimization, in some cases it is possible to hoist
key lookup into a parent loop to avoid excessive lookups.

\subsection{Galloping}
\label{subsec:galloping}

An optimization that can improve the asymptotic performance of the
generated code is to skip certain values which are guaranteed not to
produce values. For example, when computing $A \cap B$, if the
current value of $A$ is 1 and the current value of $B$ is 100, we know
that any value less than 100 cannot produce values, as $B$ would not
be present.

To exploit this, we generate a ``galloping value'' $\gallopV$ using
the current iteration state, and then advance all iterators to that
galloping value using the \texttt{skipto} method. For many types of
storage, such as a sorted list, which can use binary search or
exponential search (hence ``galloping'') in logarithmic time, this can
yield significant performance improvements.
In fact, when this method is used, the generated loop structure is the
same as leapfrog triejoin~\cite{leapfrog-triejoin}, which is (within a log factor) a
worst-case optimal join algorithm for multi-way inner joins.
This connection has been noted previously by~\citet{indexed-streams}.

Our code generation algorithm computes galloping values bottom-up from an input expression.
The structure of the galloping expression is created at compile time,
and the value is computed every loop iteration.
This is summarized in the following set of equations, where $i_N$ is
the current value of input $N$. If the value of $\gallopV$ is $-\infty$, galloping is not possible.

\vspace{-\abovedisplayskip}
{\footnotesize\begin{align*}
    \gallopV(N) &= \infty \text{, where $N$ has already dropped out} &
        \gallopV(e_1 \cap e_2) &= \max(\gallopV(e_1), \gallopV(e_2)) \\
    \gallopV(N) &= i_N \text{, where $N$ is an iterator} &
        \gallopV(e_1 \cup e_2) &= \min(\gallopV(e_1), \gallopV(e_2)) \\
    \gallopV(N) &= -\infty \text{, where $N$ can only perform lookup} &
       \gallopV(e_1 - e_2) &= \gallopV(e_1) \\
    & & \gallopV(\bar{e}) &= -\infty
\end{align*}}

\section{Evaluation}
\label{sec:evaluation}

The purpose of the abstract loop IR (ALIR) in our paper is to bring relational operations into an abstract loop form to which we can apply optimizations like fusion and that can be lowered to low-level code from first principles. Our general code generator for relational algebra has four distinct advantages, for which we provide evidence in this section:
\begin{enumerate}
    \item The code generator can generate code for any relational algebra operation, including outer joins, left joins, set differences, and non-equijoins.
    \item The code generator can generate code that is fused through two or more joins to generate worst-case optimal joins (\ref{sec:fusion}).
    \item The code generator generates nested loops that iterate hierarchically over attributes, which enables filter operations to skip over many rows at a time (\ref{sec:hierarchical}).
    \item The code generator can generate code that is portable across data structures, which allows for flexibility depending on the application (\ref{sec:datastructures}).
\end{enumerate}
In addition, we show that our code generator can generate code that is competitive with other leading database systems on the TPC-H queries they were optimized for (\Cref{sec:tpch}).

\emph{Benchmark setup.}
All benchmarks were run on dual Intel Xeon E5-2640 v4 CPUs with 251 GiB of memory. The generated C++ code was compiled using g++ 10.5.0. Compilation time is not included in the results, and queries were run multiple times to reduce the effect of caching and overhead on the results. Benchmarks are single-threaded where not otherwise indicated.

\emph{Code generation time.}
The time taken to generate C++ code from our Python DSL ranged from 0.04 to 0.14~s (geomean: 0.062~s; see Appendix B). This includes file I/O, and could be further optimized.

\subsection{Generality}
\label{sec:generality}

\begin{figure}
    \centering
    \begin{tikzpicture}[
    every node/.style={text height=1.5ex, text depth=0.25ex},
    rel/.style={draw, rectangle, minimum width=3cm},
    x=5cm, y=1.5cm, >={Stealth[length=5pt]},
    scale=0.75, transform shape
]

\begin{scope}[name prefix=q2,xshift=-10cm,yshift=2cm]
\node[rel,fill=lsqborange] (p1) at (0, 0) {person1: Person};
\node[rel,fill=lsqborange] (p2) at (1, 0) {person2: Person};
\node[rel,fill=lsqbblue] (co) at (0, -1) {Comment};
\node[rel,fill=lsqbred] (po) at (1, -1) {Post};
\draw[->] (co) -- node[above] {replyOf} (po);
\draw[->] (co) -- node[right] {hasCreator} (p1);
\draw[->] (po) -- node[right] {hasCreator} (p2);
\draw[line width=2pt] (p1) -- node[above] {knows} (p2);
\node[draw=none,yshift=-0.75cm] at (0.5, -1.25) {\textbf{Q2}};
\end{scope}
\vspace{-1em}
\begin{scope}[name prefix=q4,xshift=2cm,yshift=1.25cm]
\node[rel,fill=lsqbpink] (ta) at (-0.5, 1) {Tag};
\node[rel,fill=lsqborange] (cr) at (0.5, 1) {creator: Person};
\node[rel,fill=lsqborange] (li) at (-0.5, -1) {liker: Person};
\node[rel,fill=lsqbblue] (co) at (0.5, -1) {Comment};
\node[rel,fill=lsqbpurple] (me) at (0, 0) {Post $\cup$ Comment};
\draw[->, line width=2pt] (me) -- node[pos=0.45,left,xshift=-0.2cm] {hasTag} (ta);
\draw[->] (me) -- node[pos=0.45,right,xshift=0.2cm] {hasCreator} (cr);
\draw[->, line width=2pt] (li) -- node[pos=0.55,left,xshift=-0.2cm] {likes} (me);
\draw[->] (co) -- node[pos=0.55,right,xshift=0.2cm] {replyOf} (me);
\node[draw=none] at (0, -1.25) {\textbf{Q4}};
\end{scope}
\vspace{-1em}
\begin{scope}[name prefix=q6,xshift=-10cm,yshift=-0.5cm]
\node[rel,fill=lsqborange] (p1) at (1, -1) {person1: Person};
\node[rel,fill=lsqborange] (p2) at (0, -1) {person2: Person};
\node[rel,fill=lsqborange] (p3) at (0, -2) {person3 $\ne$ person1};
\node[rel,fill=lsqbpink] (ta) at (1, -2) {Tag};
\draw[line width=2pt] (p1) -- node[above] {knows} (p2) -- node[right] {knows} (p3);
\draw[->, line width=2pt] (p3) -- node[above] {hasInterest} (ta);
\node[draw=none] at (0.5, -2.25) {\textbf{Q6}};
\end{scope}
\vspace{-1em}
\begin{scope}[name prefix=q9,xshift=-0.5cm,yshift=-0.5cm]
\node[rel,fill=lsqborange] (p1) at (1, -1) {person1: Person};
\node[rel,fill=lsqborange] (p2) at (0, -1) {person2: Person};
\node[rel,fill=lsqborange] (p3) at (0, -2) {person3 $\ne$ person1};
\node[rel,fill=lsqbpink] (ta) at (1, -2) {Tag};
\draw[line width=2pt] (p1) -- node[above] {knows} (p2) -- node[right] {knows} (p3);
\draw[->, line width=2pt] (p3) -- node[above] {hasInterest} (ta);
\draw[line width=2pt, dashed, red] (p1) to node[above left=-0.2em] {$\textrm{!knows}$} (p3);
\node[draw=none] at (0.5, -2.25) {\textbf{Q9}};
\end{scope}
\vspace{-1em}
\end{tikzpicture}
    \caption{Selected LSQB queries. For each query, the result is the total number
    of matches. Lines and arrows represent inner joins; arrows are used when the relation is directed (not symmetric).
    Thick lines represent many-to-many relations. Red dashed lines represent negation.}
    \label{fig:lsqb-diagram}
    \vspace{-1em}
\end{figure}
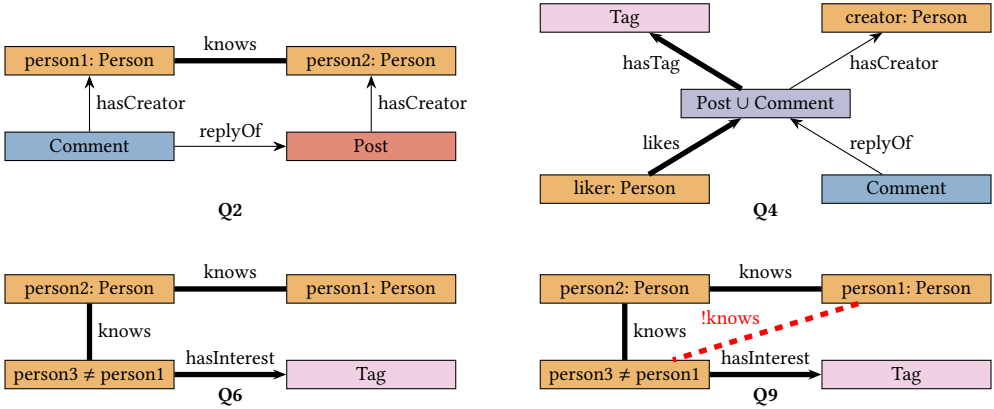

\begin{figure}
    \centering
    \input{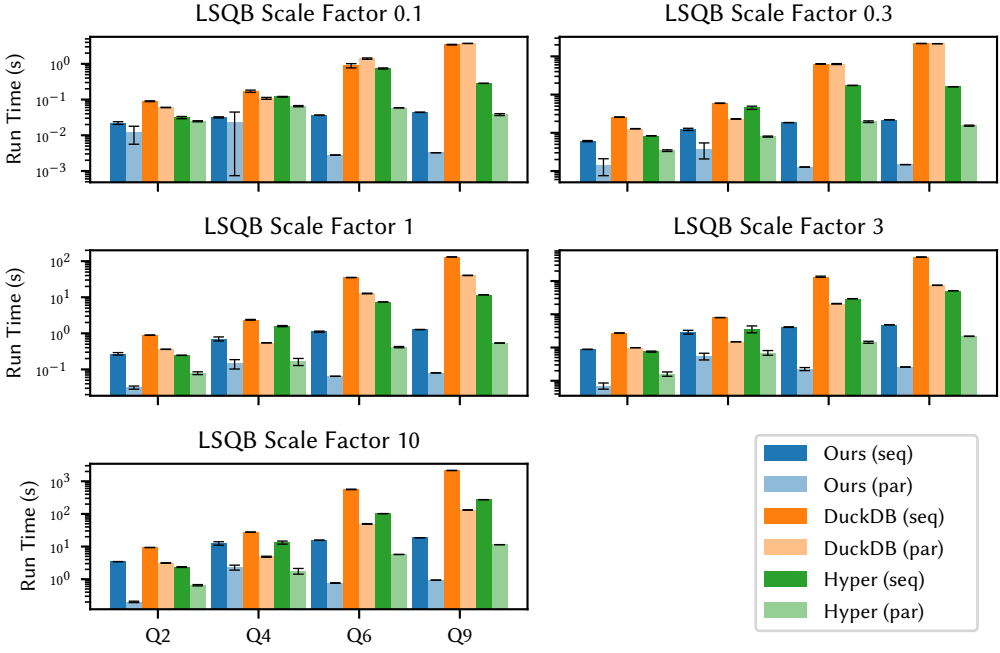}
    \vspace{-2em}\caption{LSQB results.}
    \label{fig:lsqb}
    \vspace{-2em}
\end{figure}

To demonstrate the generality of our approach, we evaluated our code generator on a diverse set of operations, against DuckDB~\cite{duckdb} and Hyper~\cite{hyper}, using the Tableau Hyper API, on four selected queries from the LSQB~\cite{lsqb} benchmark. As shown in \Cref{fig:lsqb-diagram}, the selected benchmarks have various types of operations. Our code generator is able to generate fully fused code for all four queries. The results are shown in Figure~\ref{fig:lsqb} for both sequential and parallel runtime.
The LSQB queries end with a count operation in order to not stress the client application to retrieve all of the data. However, existing database systems would typically materialize the output tuples.
In order to make the comparison fair, our code generator outputs each found tuple into a temporary buffer. If we optimize this operation away through fusion, performance increases by a factor of up to 7.5$\times$: sequential speedup is 7.5$\times$ for Q6 and 2.9$\times$ for Q9 and parallel speedup is 4.8$\times$ and 2.6$\times$. Q2 and Q4 are within variance.

For query 2, our code is slightly slower sequentially than Hyper at SF 10 (speedup: 0.68$\times$), but we achieve a 3.24$\times$ speedup in parallel. For query 4, sequential runtime is the same as Hyper, and we are slightly slower in parallel (speedup: 0.77$\times$). Queries 6 and 9 are significantly faster: Q6 has a 6.54/7.51$\times$ speedup (sequential/parallel), and Q9 has a 14.66/12.23$\times$ speedup.

We attribute the significant speedup in queries 6 and 9 to a more
efficient query plan that uses a multi-way join and hierarchical iteration
to improve query performance. The query plans generated by both Hyper and DuckDB compute a binary join between \textbf{Person\char`_knows\char`_Person} and \textbf{Person\char`_hasInterest\char`_Tag} before performing another join on \textbf{Person\char`_knows\char`_Person}. In contrast, our query plan is able to defer iterating through tags until all three people have been determined, and for Q9 we are able to use a multi-way join with the anti-join on \textbf{knows}.

\subsection{Fusion}
\label{sec:fusion}

\begin{figure}
    \begin{minipage}[b]{0.34\linewidth}\vspace{0pt}
        \centering\input{tri_results.pgf}
        \vspace{-0.3cm}\caption{Triangle query.
        Ours $\sim N^{0.99}$, DuckDB $\sim N^{1.97}$, Hyper $\sim N^{1.77}$.}
        \label{fig:tri}
        \vspace{-1em}
    \end{minipage}
    \begin{minipage}[b]{0.27\linewidth}\vspace{0pt}
        \centering
        \input{hier_results.pgf}
        \vspace{-0.3cm}\caption{Hierarchical iteration comparison.}
        \label{fig:hier}
        \vspace{-1em}
    \end{minipage}
    \begin{minipage}[b]{0.36\linewidth}\vspace{0pt}
        \centering
        \input{inter_results.pgf}
        \vspace{-0.8cm}\caption{Join comparison using intersection. Error bars represent the standard deviation of each segment. Bottom chart is normalized to fastest kernel.}
        \label{fig:inter}
        \vspace{-1em}
    \end{minipage}
    \centering
\end{figure}

To demonstrate the advantages of fusion, we implemented the
triangle-finding query~\cite{ngo-skew}:
$A(a,b) \ijoin B(b,c) \ijoin C(a,c)$,
where
$A, B, C = \{ (1, i) \mid i \in 1, \dots, n \} \cup \{ (i, 1) \mid i \in 1, \dots, n \}$,
which requires $\Theta(n^2)$ runtime for
any join-based execution strategy.
However, our approach can generate
code with optimal asymptotic efficiency $\Theta(n)$
for this query, as shown in Figure~\ref{fig:tri}.
As shown, we significantly outperform both DuckDB and Hyper,
which exhibit quadratic scaling.

\subsection{Hierarchical Iteration}
\label{sec:hierarchical}

To clearly show the performance advantage of hierarchical iteration,
we compared flat and hierarchical iteration on a simple query\footnote{
Query computed is \allowbreak
        $_{u}\agg_{\textbf{Count}(v)}(\allowbreak
            \sigma_{u.\textit{language} = \textrm{`EN'} \land
            u.\textit{created\char`_at} \textrm{ is between 1/1 and 1/4}}(\allowbreak
                \textrm{Features} \ljoin_u
                    \sigma_{v.\textit{dead\char`_account}}(\textrm{Edges} \ijoin_v \textrm{Features})
            ))$.
}, shown in
Figure~\ref{fig:hier}, on the Twitch Gamers network~\cite{rozemberczki2021twitch}.
For this query, the filter on $u$ is able to remove a large portion of edges
(selectivity 1/159.9), which allows us to skip processing them entirely
using hierarchical iteration. For this evaluation, we used column storage for
both queries; trie storage would show a greater performance difference for this
query, as the non-matching rows could be directly skipped over.

\subsection{Data Structures}
\label{sec:datastructures}

Databases support several internal data structures for different access patterns. A well-known example is sort-merge join and hash join, where most databases implement both. Sort-merge join requires both inputs to be sorted, and hash join requires one input to be stored in a hash table.

We compared the performance of different join algorithms in our system by varying the storage of the joined relations.
As shown in Figure~\ref{fig:inter}, we compared sort-merge join, sort-merge join with galloping, and hash join.
The results show the runtime of computing the intersection of two relations $A$ and $B$, where $A$ and $B$ are generated by sampling $d N$ elements from $[1, N]$, for three choices of densities $d_A:d_B$. In all cases, $N = 10{,}000{,}000$.
The ``native'' bar shows the runtime of a hash join using a C++ \texttt{unordered\char`_map}. This represents the case where a user already has the data in memory, but not in a performant data structure used by conventional database systems.

Sort-merge join has excellent performance when the input sizes are similar, but it is unable to exploit sparsity when one of the inputs is very sparse, as shown by the 0.001:0.8 results. Galloping alleviates this problem at a modest performance cost, most visible with a slight imbalance (0.4:0.6) exhibiting worst-case behavior for the cache and branch predictor.
Hash join can exploit sparsity in all cases, but it has a higher constant factor than sort-merge join and is slightly slower to build.

One application of data structure portability is the ability to integrate seamlessly with existing systems. For this example, we have added a storage description for C++ \texttt{unordered\char`_map}s with lookup capability.
This takes a total of 6 lines of Python code. As shown in Figure~\ref{fig:inter},
converting between storage formats can have a significant runtime cost, especially if the data is only used once or a few times, but in all cases the native performance is significantly worse than some other implementation. Depending on the user's requirements, they can choose whether or not to convert based on how many times the query is run for each conversion.

\subsection{TPC-H Comparisons}
\label{sec:tpch}

\begin{figure}
    \centering
    \scalebox{0.5}{\input{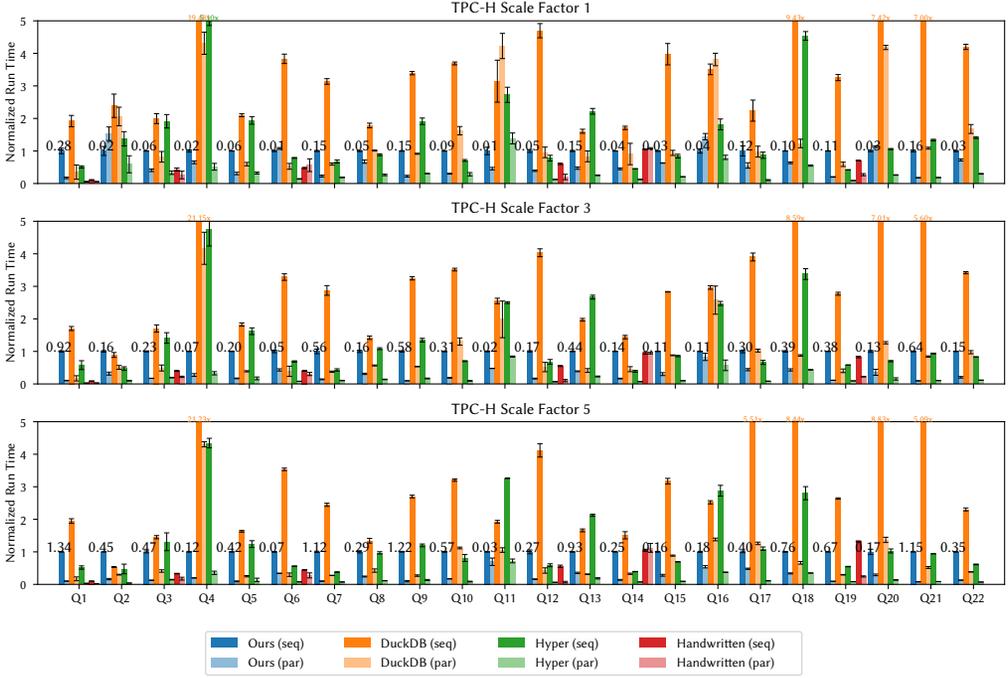}}
    \vspace{-0.5em}
    \caption{TPC-H results. All results normalized to our sequential runtime, which is labeled in seconds.}
    \label{fig:tpch}
    \vspace{-1em}
\end{figure}

Finally, we tested our performance on the TPC-H~\cite{tpch} benchmark.
In addition to DuckDB and Hyper, we also compared against hand-written
TPC-H queries~\cite{handwritten-tpch}
when available.

The ALIR for TPC-H was automatically generated and then hand-tuned; see Appendix C for details.
As shown in Figure~\ref{fig:tpch}, we outperform all databases in 10/22 of the benchmarks when run sequentially, achieving 0.378--4.343$\times$ speedup (geomean: 1.0045$\times$).
In parallel, we outperform all databases in 7/22 of the benchmarks,
achieving 0.225--1.804$\times$ speedup (geomean: 0.6114$\times$).

\section{Related Work}
\label{sec:related-work}

\paragraph{Database Management Systems}

Relational database management systems go back to System R~\cite{chamberlin1981} and Ingres~\cite{stonebraker1976}. The Volcano iterator model~\cite{volcano} was proposed for query execution, where relational operators are organized in a tree where parents request tuples from children.  The worst-case optimal join algorithms~\cite{ngo-skew,leapfrog-triejoin} showed that fusing inner joins leads to improved worst-case asymptotic complexity.

\emph{Query Compilation.}
As databases started fitting into memories and thus were no longer dominated by disk access times, researchers explored query compilation. The MonetDB system~\cite{monetdb} first demonstrated the benefits of query compilation. The HyPer system~\cite{hyper} proposed a compilation approach using LLVM that fused pipelines (operations on a single relation) for improved temporal locality. \citet{kersten2018} compares the performance of the query compilation approach to a vectorized Volcano model showing both have strengths and weaknesses, while \citet{menon2017} shows how to combine compilation with vectorization. \citet{klonatos2014} show how to design query compilers using the LMS~\cite{lms} compiler framework by optimizing certain code patterns. The EmptyHeaded system~\cite{emptyheaded} describes a compiler for the worst-case optimal join algorithm that can fuse across inner joins. Finally, \citet{indexed-streams} show how to compile the natural relational algebra (with only inner joins) on set relations with fusion and data structure portability. Our work shows how to extend natural relational algebra compilation with the large set of irregular operations supported by real-world databases---e.g., outer/left joins, non-equi joins, Cartesian products, and differences---on relations with multiset semantics, while also supporting fusion and data structure portability.

\emph{Sparse Tensor Algebra Compilation.}
Our work builds on ideas for sparse tensor algebra compilation~\cite{kjolstad2017}. Multiplications and additions in sparse tensor algebra lead to co-iteration over tensor coordinates that is similar to loops for inner and outer joins. \citet{henry2017} extended the TACO compilation model with support for non-linear operations, adding the ability to iterate over set complements of tensor coordinates. \citet{chou2018} extended TACO's coordinate tree abstraction with support for more data structures,
FiberTrees abstracts and formalizes the coordinate tree representation~\cite{fibertrees}, and \citet{indexed-streams} developed a stream-based model. Our work builds on these ideas, but shows generalize them to relational algebra with bag semantics that supports complex operations such as Cartesian products, the large family of join algorithms, filters, group-bys, and aggregation.

\section{Conclusion}
We have shown how to compile the major operations supported by real-world industry analytics query engines without giving up operator fusion or data structure portability. We hope our work can serve as a stepping stone towards query execution engines that provide full support for modern query languages, support sophisticated optimizations, and is portable across data structures.

\bibliography{main}

\clearpage
\appendix
\section{Proofs}
\lemfloor*
\begin{proof}
    Let $N$ be the set of nodes in the iteration machine.
    This is satisfied by
    \[ \floor(p) = \bigcup \{ n' \in N \mid n' \subseteq p \}. \]
    First, the set of such nodes is nonempty as
    $N \ni \{\} \subseteq p$,
    and for every predecessor $n' \subseteq p$, we have
    $n' \subseteq \floor(p) = n' \cup
    \{ n'' \in N - \{n'\} \mid n'' \subseteq p \}.$
    Finally, $\floor(p) \in N$ as $N$ is closed under union.
    Monotonicity holds by transitivity.
\end{proof}

\lemlatticedfa*
\begin{proof}
    By induction on $n$. The base case for $n = 0$ is trivial,
    as the initial state is $\top$.

    For the inductive case, it is sufficient to show that for all
    $p$ and $a$,
    \[ \floor(\floor(p) - \{a\}) = \floor(p - \{a\}). \]
    Monotonicity gives $\floor(\floor(p) - \{a\}) \subseteq \floor(p - \{a\})$.
    To show the other direction, suppose there exists some $b$ such
    that $b \in \floor(p - \{a\}) \subseteq \floor(p)$ and $b \ne a$. Then
    $b \in \floor(p) - \{a\}$, so $b \in \floor(\floor(p) - \{a\})$ as desired.
\end{proof}

\lemdfauniq*
\begin{proof}
    Consider the node $d = (u_1 \cup v_1, u_2 \cup v_2)$.
    Observe that $u_1 \cup v_1$ dominates both $u_1$ and $v_1$
    (any sequence of transitions that reaches $u_1$ or $v_1$ must pass
    through it), and similarly for $u_2 \cup v_2$.
    However, any transition that leaves $d$ must remove some $a
    \in u_1 \cup u_2 = v_1 \cup v_1 = u_1 \cup v_1 \cup u_2 \cup v_2$.
    Therefore, either $(u_1, u_2) = (v_1, v_2)$ or
    or one or both of $(u_1, u_2), (v_1, v_2)$ are not reachable.
\end{proof}

\lemctrlflow*
\begin{proof}
    By the definition of topological ordering, the next valid loop must follow any possible previous loop. For a given set of remaining inputs $I$, the corresponding loop is $\floor(I)$ by Lemma~\ref{lem:latticedfa}.
    We now show that there there can be no intermediate loop before $\floor(I)$ that satisfies its loop condition (i.e.,\ is a subset of $I$). Let $L$ be a loop before $\floor(I)$. We do not need to assume that $L$ occurs after the loop we exited.
    
    First, consider the case where $L$ is a superset of $\floor(I)$.
    By the floor function definition (Lemma~\ref{lem:floor}), $L \not\subseteq I$, so the loop condition would fail immediately.
    Now consider the case where $L$ is not a subset of $\floor(I)$. Then let $L' = L \cup \floor(I)$, which is also a valid loop by closure, and it must occur before $\floor(I)$ and $L$. Then if $L \subseteq I$, since we know $\floor(I) \subseteq I$ by definition, we must also have $L' \subseteq I$, but this reduces to the case where $L'$ is a superset of $\floor(I)$.
\end{proof}

\section{Code Generation and Compile Times}
The time for code generation and compilation are shown in the tables below. Note that this is a prototype implementation; compile time can be significantly improved by switching to a more performant backend compiler and avoiding writing the generated code to disk.

\begin{table}[H]
\begin{minipage}[t]{0.48\linewidth}
    \caption{Compilation times for TPC-H.}
    \centering
\begin{tabular}{lrrrr}\toprule
& \multicolumn{2}{c}{Code Gen. (s)} & \multicolumn{2}{c}{Compile (s)}\\\cmidrule(lr){2-3}\cmidrule(lr){4-5}
& Seq. & Par. & Seq. & Par.\\\midrule
Q1 &  0.07 &  0.06 &  2.83 &  3.27 \\
Q2 &  0.07 &  0.06 &  3.04 &  4.57 \\
Q3 &  0.06 &  0.05 &  2.41 &  3.41 \\
Q4 &  0.06 &  0.06 &  2.63 &  3.03 \\
Q5 &  0.06 &  0.06 &  2.22 &  3.58 \\
Q6 &  0.04 &  0.04 &  2.32 &  2.31 \\
Q7 &  0.06 &  0.06 &  2.43 &  3.43 \\
Q8 &  0.07 &  0.07 &  2.78 &  4.12 \\
Q9 &  0.06 &  0.06 &  2.83 &  4.26 \\
Q10 &  0.06 &  0.06 &  2.30 &  3.10 \\
Q11 &  0.06 &  0.06 &  2.08 &  3.12 \\
Q12 &  0.05 &  0.07 &  2.21 &  3.27 \\
Q13 &  0.06 &  0.06 &  2.59 &  3.91 \\
Q14 &  0.09 &  0.07 &  2.05 &  3.00 \\
Q15 &  0.10 &  0.06 &  2.62 &  3.50 \\
Q16 &  0.06 &  0.14 &  2.89 &  3.44 \\
Q17 &  0.06 &  0.07 &  2.68 &  3.05 \\
Q18 &  0.07 &  0.06 &  2.43 &  2.77 \\
Q19 &  0.06 &  0.06 &  2.42 &  3.24 \\
Q20 &  0.07 &  0.06 &  3.47 &  4.87 \\
Q21 &  0.06 &  0.07 &  2.44 &  3.33 \\
Q22 &  0.07 &  0.06 &  2.68 &  3.52 \\\bottomrule
\end{tabular}
\end{minipage}
\begin{minipage}[t]{0.48\linewidth}
\caption{Compilation times for LSQB.}
    \centering
    \begin{tabular}{lrrrr}\toprule
& \multicolumn{2}{c}{Code Gen. (s)} & \multicolumn{2}{c}{Compile (s)}\\\cmidrule(lr){2-3}\cmidrule(lr){4-5}
& Seq. & Par. & Seq. & Par.\\\midrule
Q2 &  0.06 &  0.07 &  1.61 &  2.10 \\
Q4 &  0.06 &  0.06 &  2.67 &  2.39 \\
Q6 &  0.07 &  0.06 &  1.59 &  2.30 \\
Q9 &  0.06 &  0.06 &  1.93 &  1.95 \\\bottomrule
\end{tabular}
\end{minipage}
\end{table}

\section{Automatic Query Planning and ALIR Generation}
In this section, we give an overview of our rudimentary automatic query planner, which we used to generate initial implementations of TPC-H queries. After this, we manually modified 13 of these implementations (2, 6, 8, 9, 10, 11, 13, 15, 16, 17, 18, 20, 22) for better performance; the remaining 9 queries were not modified.

The query planner operates directly on the SQL input. First, the input is parsed and analyzed, and subqueries are extracted. For each query, a suitable join ordering is found heuristically, prioritizing larger tables first. For each table in the join ordering, each attribute in the primary key is inserted as a nested loop. Tables are transformed into primary-key lookups when possible, and loops for joined tables are merged. For example, for the following query (taken from Q3):
\begin{align*}\footnotesize
    &\textbf{SELECT } \dots \textbf{ FROM } \textrm{customer, orders, lineitem} \\[-1ex]
    &\quad\quad \textbf{WHERE } \dots \textit{ c\_custkey} = \textit{o\_custkey} \textbf{ AND } \textit{l\_orderkey} = \textit{o\_orderkey} \dots
\end{align*}
the loops generated (before merging) are:
\begin{algorithmic}\footnotesize
    \ALIRFor{$o \in \textrm{lineitem}.\textit{orderkey}$}
        \ALIRFor{$l \in \textrm{lineitem}.\textit{linenumber}$}
            \ALIRFor{$o' \in \textrm{order}.\textit{orderkey}$}
                \ALIRFor{$c \in \textrm{customer}.\textit{custkey}$}
                    \State \dots
                \EndALIRFor
            \EndALIRFor
        \EndALIRFor
    \EndALIRFor
\end{algorithmic}
Based on the join conditions, the loops for $o'$ and $c$ can both be merged into the loop for $o$, which results in the following loops:
\begin{algorithmic}\footnotesize
    \ALIRFor{$o \in \textrm{lineitem}.\textit{orderkey} \cap \textrm{order}.\textit{orderkey} \cap \textrm{customer}$}
        \ALIRFor{$l \in \textrm{lineitem}.\textit{linenumber}$}
            \State \dots
        \EndALIRFor
    \EndALIRFor
\end{algorithmic}
Here, \textrm{customer} is transformed into a primary-key lookup based on the value of $\textrm{order}.\textit{custkey}$. This is not indicated in the ALIR, but rather in the storage description.

After the loop ordering is decided, predicates are inserted as virtual relations in the outermost applicable loop. For this example, conditions depending on \textrm{customer} or \textrm{orders} are appended to the $o$ loop, and conditions depending on \textrm{lineitem} are appended to the $l$ loop.

Next, body statements to either output a tuple or update an aggregation value are inserted in the innermost loop. Finally, code is generated to sort the results as necessary and produce output tuples based on the aggregation results.

To generate parallel code, we parallelize the outermost loop of each query using OpenMP. Each thread begins by tiling the first iterator relation in the outer loop domain into $N$ tiles, where $N$ is the number of threads. After this, the tile boundaries must be adjusted so that it does not split a node in the coordinate tree; the loop must either handle all tuples with the corresponding attribute or none, as partial handling could result in incorrect multiset semantics. This is shown in the following diagram, representing a tiling of 6 values on 2 threads:
\[ [1\quad1\quad2]\quad[2\quad2\quad2] \implies [1\quad1]\quad[2\quad2\quad2\quad2]. \]
An even tiling, shown on the left, splits the value 2, so tile boundaries are moved so that this does not occur, shown on the right. Given this, the remaining iterators are tiled based on the tiles for the first relation. Each thread then computes its local result, and the local results are aggregated to form the final result.

The initial implementations produced by this query planner are sufficient to generate working code for TPC-H queries, but in many cases they generate suboptimal code. In particular, the query planner does not attempt to optimize subqueries, which we found left significant performance gains to be realized. Also, these implementations can lead to severe bottlenecks in parallel execution. As such, we manually modified several benchmarks to improve performance.

\end{document}